\documentclass[aps,prb,twocolumn,floatfix,showpacs,groupedaddress,superscriptaddress]{revtex4-1}

\usepackage{graphicx}  
\usepackage{dcolumn}   
\usepackage{braket}
\usepackage[mathscr]{euscript}
\usepackage{amsbsy}
\usepackage{bm}        
\usepackage{amssymb}
\usepackage[T1]{fontenc} 
\usepackage{amsmath}
\usepackage{xcolor}
\usepackage[caption=false,labelformat=simple]{subfig}

\begin{document}

\title{Magneto-optical properties of a semi-Dirac nanoribbon in the terahertz  frequency regime}
\author{Priyanka Sinha}
\email{sinhapriyanka2016@iitg.ac.in}
\affiliation{Department of Physics, Indian Institute of Technology Guwahati\\ Guwahati-781039, Assam, India}
\author{Shuichi Murakami}
\email{murakami@stat.phys.titech.ac.jp}
\affiliation{Department of Physics, Tokyo Institute of Technology\\
2-12-1 Ookayama, Meguro-ku, Tokyo 152-8551, Japan}
\author{Saurabh Basu}
\email{saurabh@iitg.ac.in}
\affiliation{Department of Physics, Indian Institute of Technology Guwahati\\ Guwahati-781039, Assam, India}
\date{\today}
\begin{abstract}
We study magneto-optical (MO) properties of a semi-Dirac nanoribbon in presence of a perpendicular magnetic field using Kernel Polynomial Method (KPM) based on the Keldysh formalism in the experimental (terahertz frequency) regime. For comparison, we have also included results for the Dirac systems as well, so that the interplay of the band structure deformation and MO conductivities can be studied. We have found that the MO conductivity shows features in the semi-Dirac system that are quite distinct from the Dirac case near the ultra-violet and the visible regimes. The real parts of the longitudinal conductivities, namely Re($\sigma_{xx}$) and Re($\sigma_{yy}$) (which are different in the semi-Dirac case, as opposed to a Dirac one) present a series of resonance peaks as a function of the incident photon energy. We have also found that the absorption peaks corresponding to the $y$-direction are larger (roughly one order of magnitude) than those corresponding to the $x$-direction for the semi-Dirac case. In the case of the Hall conductivity, that is, Re($\sigma_{xy}$), there are extra peaks in the spectra compared to the Dirac case which originate from the distinct optical transitions of the carriers from one Landau level to another. We have also explored how the carrier concentration influences the MO conductivities. In the semi-Dirac case, there is the emergence of additional peaks yet again in the absorption spectrum underscoring the presence of an asymmetric dispersion compared to the Dirac case. Further, we have explored the interplay between the polarization of the incident beam and the features of the absorption spectra which can be probed in experiments. Finally, we evaluate the MO activity of the medium by computing the Faraday rotation angle, $\theta_{F}$. 
\end{abstract}
\maketitle

\section{Introduction}
Over the past few years, the discovery of graphene \cite{wallace,neto,geim} as well as other two-dimensional (2D) materials, such as silicene \cite{yan}, phosphorene \cite{du,kong}, MoS$_2$ \cite{lebe,mak,tahir1}, $8$-pmmn borophene \cite{lozovik} etc. have enriched our knowledge on many of the experimental and theoretical aspects \cite{cava,fuch,adam} of these materials owing to their low-energy physics being governed by massless Dirac particles. The spectrum of a massless Dirac particle has two cones, the so-called “Dirac cones” in the vicinity of two non-equivalent points $K_1$ and $K_2$ in the Brillouin zone (BZ). Recently, a close variant of the 2D Dirac materials termed as the semi-Dirac materials have been discovered. In a tight-binding model for a 2D Dirac system, such as graphene, consider that one of the three nearest-neighbor (nn) hopping energies is tuned (say, $t_2$) with respect to the other two (say, $t$), the two Dirac points with opposite chiralities move in the BZ. Eventually, when $t_2$ becomes equal to $2t$, the two cones merge into one at the so-called semi-Dirac point. Such evidences of merging of the Dirac points has been observed in fermionic cold atomic gases loaded into an optical lattice \cite{tarruell}. Such materials possess unique band dispersion which simultaneously shows massless Dirac (linear) along one direction and massive fermionic (quadratic) features in other directions, leading to an anisotropic electronic dispersion\cite{dietl,pickett}. Such dispersion is found in phosphorene under pressure and doping \cite{rodin,guan}, electric fields \cite{rudenko,dut}, TiO$_2$/VO$_2$ superlattices \cite{pickett, pardo1, pardo2}, graphene under deformation \cite{mon}, and BEDT-TTF$_{2}$I$_{3}$ salt under pressure \cite{konno,pichon}. Several properties of the semi-Dirac system have been discussed in literature\cite{pickett,ban} including the effect of the merging Dirac points on the emergence of a Chern insulating state\cite{kush}, the presence of Chern insulating state including spin-orbit coupling\cite{duan}, the topological phase transition driven by disorder\cite{kush2}, the Floquet topological transition in graphene by an ac electric field \cite{pla}, and the orbital susceptibility in dice lattice\cite{ra} etc.\par
Further, the behavior of the Dirac fermions in graphene has been studied in the presence of an external magnetic field, which facilitated the realization of half-integer quantum Hall effect at room temperature\cite{novo,vp}. When an external magnetic field $B$ is applied perpendicularly to the plane of the sample, the energy spectrum transforms into discrete Landau levels and the level energies, $E_n$ takes the form, $E_n\propto\sqrt{|n|B}$, where $B$ is the magnetic field and $n$ denotes Landau level indices \cite{sadowski1,sadowski2}. The dependence of the Landau level energy deviates from $\sqrt{B}$ for semi-Dirac systems\cite{pickett} and it varies as $\big((|n+\frac{1}{2}|)B\big)^{2/3}$. Very recently, the study of Landau levels has been done extensively where the quantization of the conductance plateaus shows the integer quantum Hall effect for semi-Dirac system\cite{sinha3}. \par
Usage of optical probe for the Dirac materials has gathered momentum on a parallel ground in recent years. The vector potential of the incident photons couple to the band electrons via Peierls' coupling. The situation becomes more complicated in the presence of an external magnetic field where the kinetic energy of the carriers transforms into macroscopically degenerate Landau levels. The MO transport properties of these materials are gradually studied in the linear regime using the Kubo formula. However, evaluating the effects of deformation of the band structure on the transitions induced by optical means for the carriers from one Landau level to another is a harder task. In the following, we present a systematic exploration of the MO transport for a semi-Dirac system in the visible frequency range.\par  
Terahertz frequency ($0.1$~THz to $10$~THz) is extensively used in emerging fields such as spectroscopy, communication, and imaging \cite{koch,antes}. Many interesting MO phenomena, such as giant Faraday rotation \cite{marel}, gate-tunable magneto-plasmons \cite{xia}, non-linear transport driven by the light radiation\cite{gani} have been discovered with graphene exposed to radiation at the terahertz frequencies. At these particular frequencies, graphene supports the propagation of  plasmon-polaritons\cite{shen,lez,polini} that can be tuned by the external gate voltage. Parallelly, it helps for the basic studies of the interaction of radiation with the matter at nanoscale dimensions\cite{lund}. Here, we show the emergence of strong magneto-absorption in the terahertz regime where the absorption peaks are well-observed. Also, optical conductivity has always yielded very useful information on the electronic transitions in presence of a time-varying driving field. This facilitates observing frequency-dependent (ac) conductivity. The real part of the longitudinal optical conductivity gives information on the absorption properties as a function of photon energy, while the imaginary part contains the information about the transmission. In the case of the optical conductivity, a photon can induce a transition between these Landau levels, and the optical frequency matches with the energy level difference of the Landau levels \cite{gusy} resulting in the absorption peaks. The characteristics of the band dispersion and the energy gap can be found from these absorption lines in the experiments \cite{tung,sadowski1,post}. In the presence of an external magnetic field, similar information emerges for the system, except that now the energy spectrum comprises Landau levels, as opposed to single-particle energies. MO properties of graphene have been studied both theoretically and experimentally and the result shows good agreement between the theoretical findings and the experiments \cite{gusy1,hao}. Also, MO properties of topological insulators \cite{tse} and other two-dimensional materials, such as MoS$_2$ \cite{chu} and silicene \cite{nicol}, phosphorene \cite{tahir2} have been studied. A recent study on the optical conductivity in three-dimensional materials has provided valuable information on quasicrystals, as well as on Dirac \cite{wang} and Weyl semimetals \cite{ashby,shao,aragon}. It is well known that the Landau level spacings are proportional to the magnetic field, $B$ when a quadratic term in Hamiltonian is alone there (for example, a 2D electron gas), while the linear term alone yields spacings as the square root of $B$ (for example, graphene) which are quite different. This has important implications for the optical absorption in a situation when both terms are present. Very few studies on optical conductivity in semi-Dirac materials have been done and hence the anisotropy in the spectra corresponding to the different planar direction ($x$ and $y$) remain largely unexplored\cite{sinner,car,ales}. A detailed and systematic study is indeed needed on the MO transport properties of the semi-Dirac materials to enhance our understanding of these systems.\par   
In this paper, we study the MO conductivity in a perpendicular magnetic field of a semi-Dirac system using a tight-binding Hamiltonian on a honeycomb lattice. We use a numerical tool based on the Keldysh formalism for large-scale calculations for any realistic system\cite{simao}. First, we study the optical conductivity in presence of a magnetic field of a semi-Dirac nanoribbon. Hence, by applying a perpendicular magnetic field, we look for the possible optical transitions that occur between the Landau levels by absorption of photons. We further calculate the longitudinal as well as the (optical) Hall conductivities as a function of the photon energy for moderate as well as very high values of the magnetic field in the terahertz regime. We also explore the effects of the carrier concentration of the Landau levels on the optical spectra by varying the chemical potential. Further, we report the conductivity for a different polarization of the incident light, such as circularly polarized radiation. Moreover, we study the effects of Faraday rotation for the semi-Dirac as well as the Dirac systems. The Faraday rotation occurs in an active MO medium where the plane of polarization of the transmitted radiation is rotated with respect to that of the incident radiation. The effect is characterized by the angle by which the plane of polarization is rotated and is called the Faraday angle\cite{marel,fer,yoo} which we compute for the semi-Dirac and the Dirac cases. \par 
The paper is organized as follows. The Keldysh formalism is described in sec.~\ref{II}. In sec.~\ref{III}, we provide numerical results for the optical conductivity in presence of a perpendicular magnetic field ($B \neq 0$). To illustrate the MO conductivity, we show the Landau level spectra and the optical transitions in sec.~\ref{A}. The transport properties are investigated by computing the Hall and the longitudinal conductivities in sec.~\ref{A}. In sec.~\ref{B}, we observe the impact of chemical potential on the real part of the longitudinal conductivities ($\sigma_{xx}$ and $\sigma_{yy}$). We see the effects of using a circularly polarized light in sec.~\ref{C}. Sec.~\ref{D} includes a brief discussion on the Faraday effect. Finally, we conclude our results in sec.~\ref{IV}.
\section{Keldysh formalism}\label{II}
To obtain the MO conductivity we use a general perturbation method, known as the Keldysh formalism \cite{keldysh}, which describes the quantum mechanical time evolution of non-equilibrium and even interacting systems at finite temperatures. A few relevant quantities that we needed are the time-ordered ($T$), anti-time-ordered ($\tilde{T}$), lesser ($G^{<}$) and greater ($G^{>}$) Green's functions which are defined as,
\begin{equation}
iG^{T}_{ab} (t,t')= \Bigg<T\Big[c_{a}(t) c^{\dagger}_{b}(t')\Big]\Bigg> \nonumber
\end{equation} \vspace{-0.2 cm}
\begin{equation}
iG^{<}_{ab} (t,t')= -\Bigg<c^{\dagger}_{b}(t') c_{a}(t)\Bigg>  \nonumber
\end{equation}  \vspace{-0.2 cm}
\begin{equation}
iG^{>}_{ab} (t,t')= \Bigg<c_{a}(t) c^{\dagger}_{b}(t')\Bigg>    \nonumber
\end{equation}  \vspace{-0.2 cm}
\begin{equation}
iG^{\tilde{T}}_{ab} (t,t')= \Bigg<\tilde{T}\Big[c_{a}(t) c^{\dagger}_{b}(t')\Big]\Bigg>.
\label{eq:1} 
\end{equation}
The time-ordering operator and the anti-time-ordering operator are denoted by $T$ and ${\tilde{T}}$. The creation and the annihilation operators are in the Heisenberg picture and the labels $a$ and $b$ denote the indices for the single-particle states. The retarded and the advanced Green's functions can be written with the combination of the Green's functions in Eq.~(\ref{eq:1}) as,
\begin{equation}
G^{R}=G^{T}-G^{<}
\label{eq:5}
\end{equation} \vspace{-0.5 cm}
\begin{equation}
G^{A}=-G^{\tilde{T}}+G^{<}.
\label{eq:6}
\end{equation}
The tight-binding Hamiltonian can be expressed as, 
\begin{equation}
H_{0}=\sum_{{\bm {R_{i}}},{\bm R_{j}}} \sum_{\sigma_{1},\sigma_{2}} t_{\sigma_{1}\sigma_{2}}({\bm R_{i}},{\bm R_{j}})c^{\dagger}_{\sigma_{1}}({\bm R_{i}})c_{\sigma_{2}}({\bm R_{j}}),
\label{eq:7}
\end{equation}
where the operator $c^{\dagger}_{\sigma_{1}}$({${\bm R_i}$) creates an electron in the carbon atoms at lattice site ${\bm R_{i}}$, whereas $c_{\sigma_{2}}$(${\bm R_j}$) annihilates an electron at lattice site ${\bm R_{j}}$ with $t$ as connecting the nearest-neighbors. We have performed all our numerical calculations by using $t=2.8$ eV which corresponds to electron hopping in graphene. In a 2D Dirac system, this value may be different. The $\sigma_{1}$ and $\sigma_{2}$ are the orbitals degrees of freedom. The electromagnetic field can be introduced through Peierls' substitution as,
\begin{equation}
t_{\sigma_{1}\sigma_{2}}({\bm R_{i}},{\bm R_{j}})\rightarrow e^{\frac{-ie}{\hbar} \int^{\bm{R_{i}}}_{\bm{R_{j}}} {\bm A}({\bm r}^\prime,t).d{\bm{r^{\prime}}}} ~t_{\sigma_{1}\sigma_{2}}({\bm R_{i}},{\bm R_{j}}).
\label{eq:8}
\end{equation}
The following vector potential can be used to introduce both a static magnetic field and a uniform electric field,
\begin{equation}
{\bm A}({\bm r},t)={\bm A}_{1}({\bm{r}})+{\bm A}_{2}(t).
\end{equation}
The electric and magnetic fields are obtained via ${\bm E}(t)=-\partial_{t}{\bm A}_{2}(t)$ and ${\bm B(r)}={\bm {\nabla}}\times{\bm A}_{1}({\bm r)}$. Accordingly, $t_{\sigma_{1}\sigma_{2}}({\bm R_{i}},{\bm R_{j}})$ gets modified by the introduction of the magnetic field only. The many-particle time-dependent Hamiltonian can be described by
\begin{equation}
H(t)=H_{0}+H_{ext}(t),
\end{equation}
where $H_{0}$ is an unperturbed Hamiltonian and $H_{ext}(t)$ is the time-dependent external perturbation. The exponential in Eq.~(\ref{eq:8}) can be expanded, which results in an infinite series of operators for the full Hamiltonian, $H(t)$ as,
\begin{equation}
H(t)=H_{0}+eA^{\alpha}(t)\hat{h}^{\alpha}+\frac{1}{2!}e^{2}A^{\alpha}(t)A^{\beta}(t)\hat{h}^{\alpha\beta}+\dotsb.
\end{equation}
From the above equation, we can write the $H_{ext}(t)$ as,
\begin{equation}
H_{ext}(t)=eA^{\alpha}(t)\hat{h}^{\alpha}+\frac{1}{2!}e^{2}A^{\alpha}(t)A^{\beta}\hat{h}^{\alpha\beta}+\dotsb.
\end{equation}
Now, we are defining $V(t)=(i\hbar)^{-1}H_{ext}$ and $A(t)=\int^{\infty}_{-\infty} \frac{d\omega}{2\pi} \tilde{A}(\omega) e^{-i\omega t}$. 
After a Fourier transform we get (dropping the spatial dependence),
\begin{align}
\tilde{V}(\omega)= & \frac{e}{i\hbar}\hat{h}^{\alpha} \tilde{A}^\alpha(\omega)+\frac{e^2}{i\hbar}\frac{\hat{h}^{\alpha\beta}}{2!} \int\frac{d\omega'}{2\pi}\int\frac{d\omega''}{2\pi} \times\tilde{A}^{\alpha}(\omega') \nonumber
\\
&
\tilde{A}^{\beta}(\omega'')2\pi\delta (\omega'+\omega'' -\omega)+\dotsb,
\label{eq:9}
\end{align}
where $\hat{h}^{\alpha}=\frac{1}{i\hbar}[r^{\alpha},H]$ and $\hat{h}^{\alpha\beta}=\frac{1}{(i\hbar)^2}[r^{\alpha},[r^{\beta},H]]$. Now we define in a general sense,
\begin{equation}
\hat{h}^{\alpha_1\dotsb\alpha_n}=\frac{1}{(i\hbar)^n}[\hat{r}^{\alpha_1},\dotsb[\hat{r}^{\alpha_n},H]],
\label{eq:10}
\end{equation}
where $\hat{h}^\alpha$ is the single-particle velocity operator at the first-order and $\hat{\bm {r}}$ is the position operator. In the presence of periodic boundary conditions, the position operator is ill-defined, but its commutator with the Hamiltonian continues to be a well-defined quantity. In the real space, this commutator yields the matrix element of the Hamiltonian connecting the two sites $i$ and $j$ multiplied by the distance vector, ${\bm d}_{ij}$ between them. ${\bm d}_{ij}$ will be a well defined quantity in case of a periodic boundary condition if we define this quantity as the distance between the neighbors, instead of the difference between the two positions. Hence, $\hat{h}$ operators can be defined in position space by multiplying the Hamiltonian matrix elements with the required product of the difference vectors. The current operator can be calculated from the Hamiltonian, via $\hat{J}^{\alpha}=-\frac{1}{\Omega}\Big(\frac{\partial H}{\partial A^{\alpha}}\Big)$ (where $\Omega$ denotes the volume of the sample). $\hat{J}^{\alpha}$ also follows a series expansion due to the presence of an infinite number of ${\bm A}(t)$ terms in presence of an external perturbation, namely, 
\begin{equation}
\hat{J}^{\alpha}(t)=-\frac{e}{\Omega}\Bigg(\hat{h}^\alpha+e\hat{h}^{\alpha\beta}A^\beta(t)+\frac{e^2}{2!}\hat{h}^{\alpha\beta\gamma}A^{\beta}(t)A^{\gamma}(t)+\dotsb\Bigg)
\label{eq:11}
\end{equation}
and the first-order optical conductivity is found to be\cite{simao},
\begin{align}
& \sigma^{\alpha\beta}(\omega)= \frac{ie^2}{\Omega\omega}\int_{-\infty}^{\infty}d\epsilon f(\epsilon)\textnormal{Tr}\Bigg[\hat{h}^{\alpha\beta}\delta(\epsilon-H_0)+\frac{1}{\hbar}\hat{h}^\alpha \nonumber
\\
&
 \textsl{g}^{R}(\epsilon/\hbar+\omega)\hat{h}^\beta\delta(\epsilon-H_0)+\frac{1}{\hbar}\hat{h}^{\alpha}\delta(\epsilon-H_0)\hat{h}^\beta \textsl{g}^{A}(\epsilon/\hbar-\omega)\Bigg].
\label{eq:12}
\end{align}
The retarded and advanced Green's functions, Dirac deltas and the generalized velocity operators are written in the position basis which is expanded in a truncated series of Chebyshev polynomials\cite{viana}. The details are given in the appendix.
\section{Results}\label{III}
To clearly explain our system, a schematic honeycomb lattice geometry of a semi-Dirac ribbon is shown in Fig.~\ref{fig:sheet}, where one of the three nearest-neighbor hoppings is modified by $t_2$ and is shown by a pink line. We have considered  primitive vectors between the unit cells as, $\bm{a}_{1}=a(\sqrt{3},0)$ and $\bm{a}_{2}=a(\frac{\sqrt{3}}{2}, \frac{3}{2})$, $a$ being the distance between two consecutive carbon atoms. Also the nearest-neighbor vectors in real space are defined by $\vec{\delta}_{1} = \big(0, a\big)$; $\vec{\delta}_{2} = \Big(\frac{\sqrt{3}a}{2}, -\frac{a}{2}\Big)$ and $\vec{\delta}_{3} = \Big(-\frac{\sqrt{3}a}{2}, -\frac{a}{2}\Big)$ as shown in Fig.~\ref{fig:sheet}. Hence, the area of the unit cell is $\Omega_{c}=\frac{3\sqrt3}{2}a^2$. The tight-binding Hamiltonian considering three nearest-neighbor hopping,
\begin{align}
 H= -\sum_{\langle{ij}\rangle} (t_{ij}c^{\dagger}_{i} c_{j} +h.c.),
 \label{ham}
\end{align}
where $c^{\dagger}_{i}$($c_{j}$) creates(annihilates) an electron on sublattice $A$($B$). $t_{ij}$ is the nearest-neighbor hopping amplitude.
\begin{widetext}
The tight-binding dispersion for a semi-Dirac system can be written as,
\begin{align}
E(k)= 
\pm \sqrt{2t^2 + t_2^2 +2t^2\cos \sqrt{3}k_xa + 4t_2t \cos(3k_ya/2) \cos(\sqrt3k_xa/2)}.
\label{eq.3}
\end{align}
\end{widetext}
In Eq.~(\ref{eq.3}) when $t_2=2t$, the band structure shows an anisotropic dispersion being quadratic (non-relativistic) along the $k_x$ direction and linear (relativistic) along the $k_y$ direction as shown in Figs.~\ref{fig:kx} and~\ref{fig:ky} respectively, justifying the special feature of the semi-Dirac dispersion.
Also, the single-particle low-energy Hamiltonian based on the tight-binding model of a semi-Dirac system can be written as,
\begin{equation}
{\it{H}}=v_F p_y \sigma_y + \frac{p^2_x \sigma_x}{2m^*},
\label{lw_ham}
\end{equation}
where the momenta along the $x$ and the $y$ directions are denoted by $p_x$ and $p_y$ respectively. The Fermi velocity along the $p_y$ direction is expressed as $v_F= {3ta}/{\hbar}$. The effective mass, $m^*$ that corresponds to the parabolic dispersion along $p_x$ is $m^*=2\hbar/3ta^2$. Here we set $a=1$. The Pauli spin matrices are $\sigma_x$ and $\sigma_y$ in the pseudospin space. The dispersion relation corresponding to Eq.~(\ref{lw_ham}) can be obtained as (without a constant shift in energy),
\begin{equation}
E=\pm \sqrt{(\hbar v_{F}k_{y})^2+\bigg(\frac{\hbar^2k_{x}^2}{2m^*}\bigg)^2}, 
\label{eq.2}
\end{equation}
where $\pm$ stands for the conduction and valence band respectively. It can be noted from Eq.~(\ref{eq.2}) that the dispersion is linear along $y$-direction, whereas the dispersion along the $x$-direction is quadratic.\par
We have taken a ribbon of size 6144 $\times$ 6144, that is, the number of unit cells in each of the $x$ and $y$ directions is 6144. We further use periodic boundary conditions in our calculations. The modified hopping parameter $t_2$ is varied from $t$ to $2t$. The convergence of the peaks depends on the number of  Chebyshev moments, $M$. For reasonable accuracy, we choose a large number of Chebyshev moments, $M=4096$ \cite{rappoprt}. The value of the magnetic field $B$ is taken 400T for all the purposes. It may be noted that the value of $B$ chosen here is very high, however, computations with a lower $B$ demands a larger size of the nanoribbon. Thus a compromise is made between the system size and value of $B$ for keeping our computations numerically feasible. Nevertheless, in some situations, we have used more moderate values of $B$, namely $B=100$T. We have also checked that the value of $M$ used by us serves our purpose. However, the system size and the number of the Chebyshev moments can further be enhanced in order to minimize the fluctuations and to achieve greater accuracy.\par
\begin{figure}[!ht!]
\centering
\subfloat[]{\includegraphics[width=0.48\textwidth]{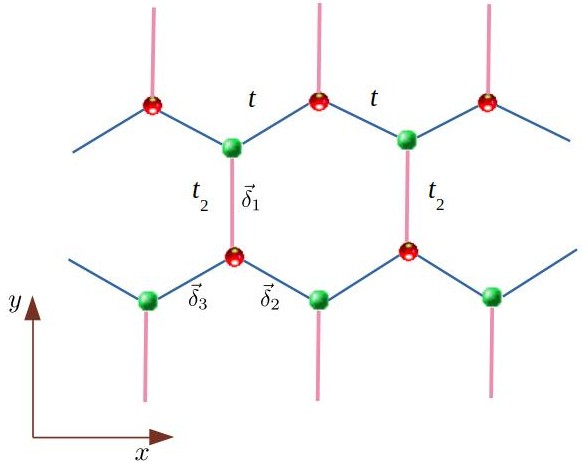}\label{fig:sheet}}\\
\subfloat[]{\includegraphics[trim=0 -20 0 0,clip,width=0.243\textwidth]{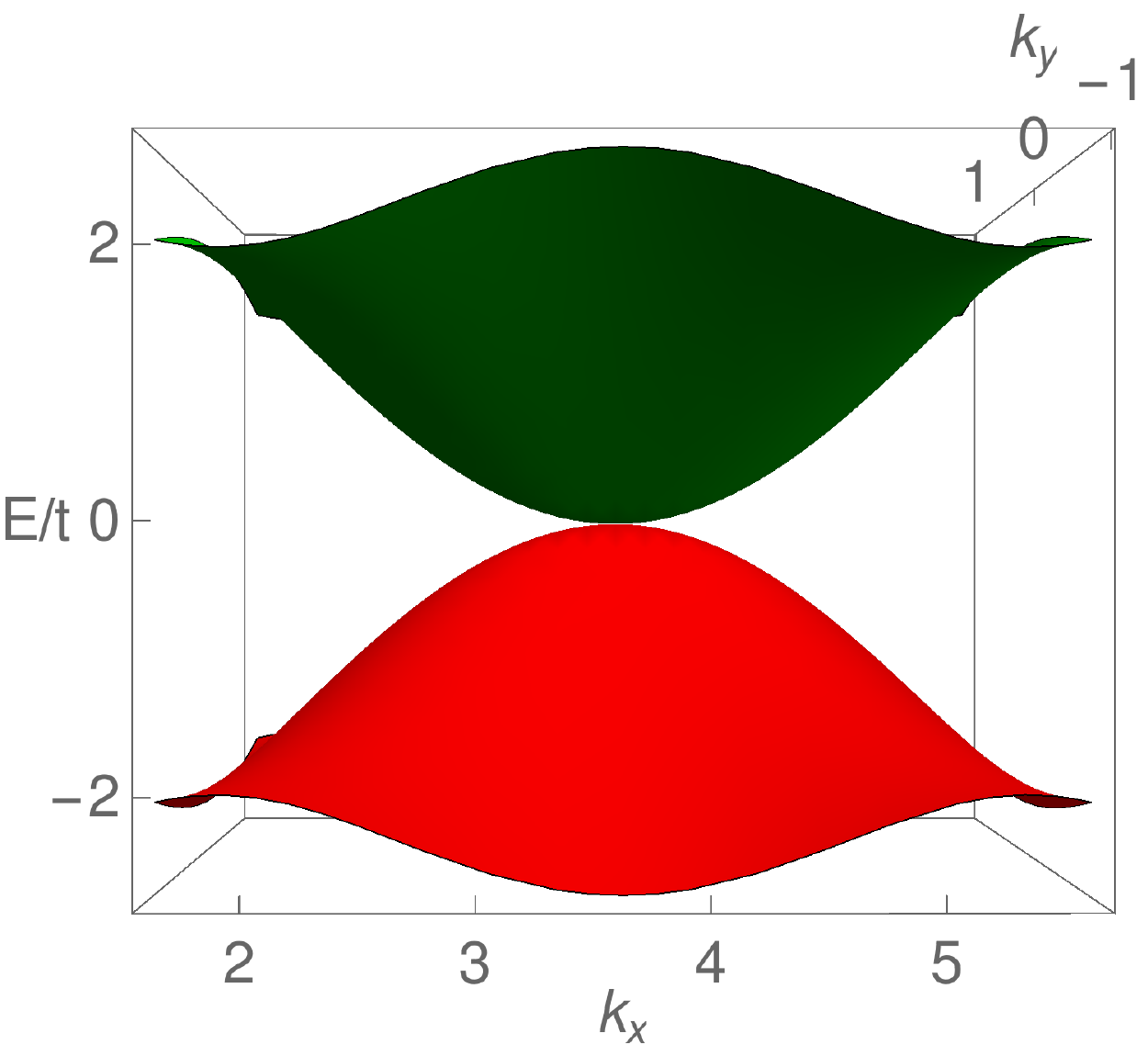}\label{fig:kx}}
\subfloat[]{\includegraphics[trim=0 0 0 0,clip,width=0.243\textwidth]{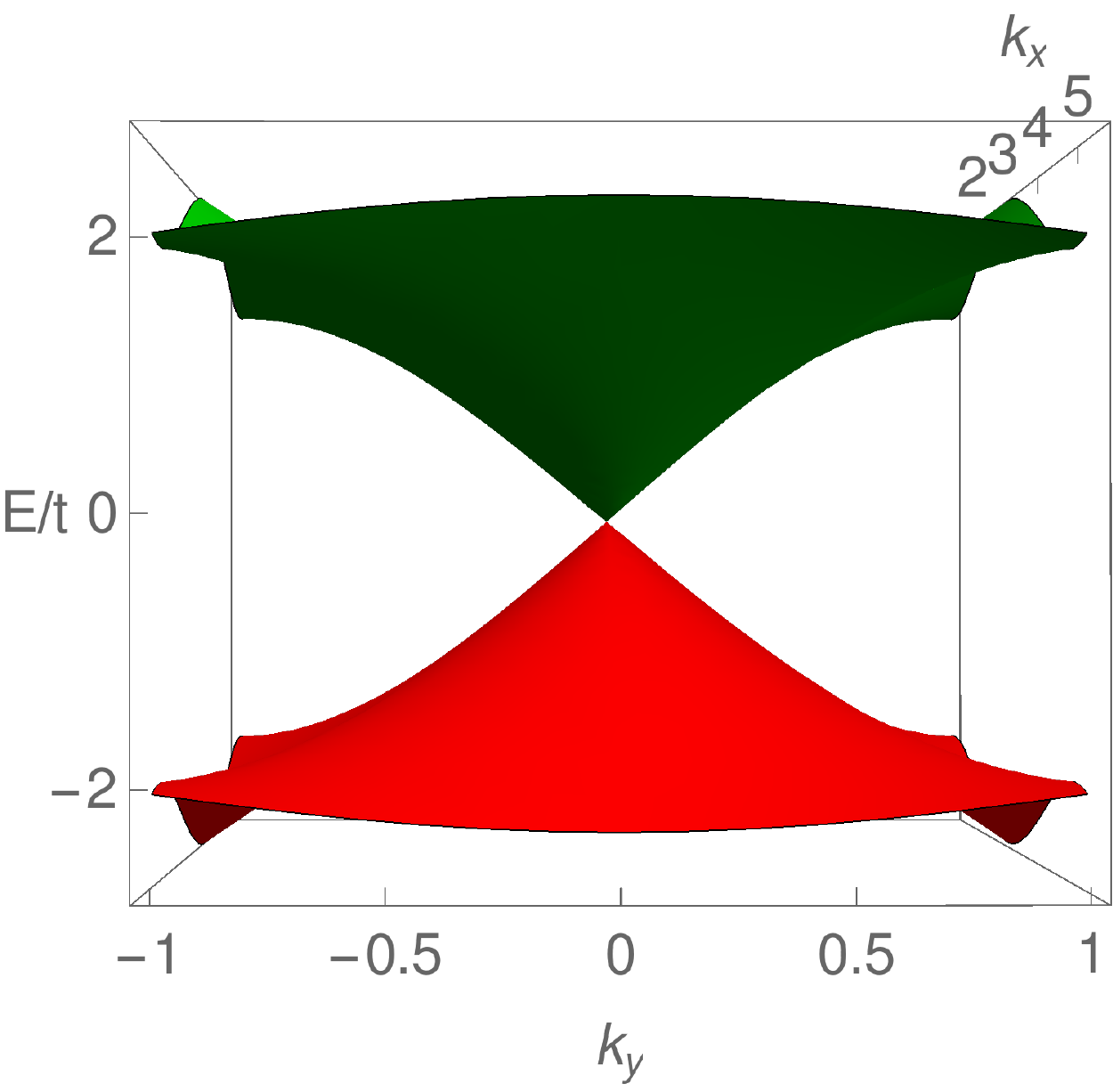}\label{fig:ky}}
\caption{(Color online) (a) A schematic sketch of the lattice geometry of a semi-Dirac system is shown with different hopping parameters $t$ (denoted by blue line) and $t_2$ (denoted by pink line). The planar directions are indicated by the $x$-$y$ axis. The nearest-neighbor vectors, $\vec{\delta}_{i}$ ($i =$ 1, 2, 3) are mentioned in the text. Carbon atoms belonging to the two sublattices are denoted by red and green colors. The anisotropic band dispersion of a tight-binding semi-Dirac system is shown along (b) the $k_x$ and (c) the $k_y$ direction.}
\label{fig:lattice}
\end{figure}
\begin{figure*}[!ht!]
\begin{center}
\subfloat{\includegraphics[width=0.4\textwidth]{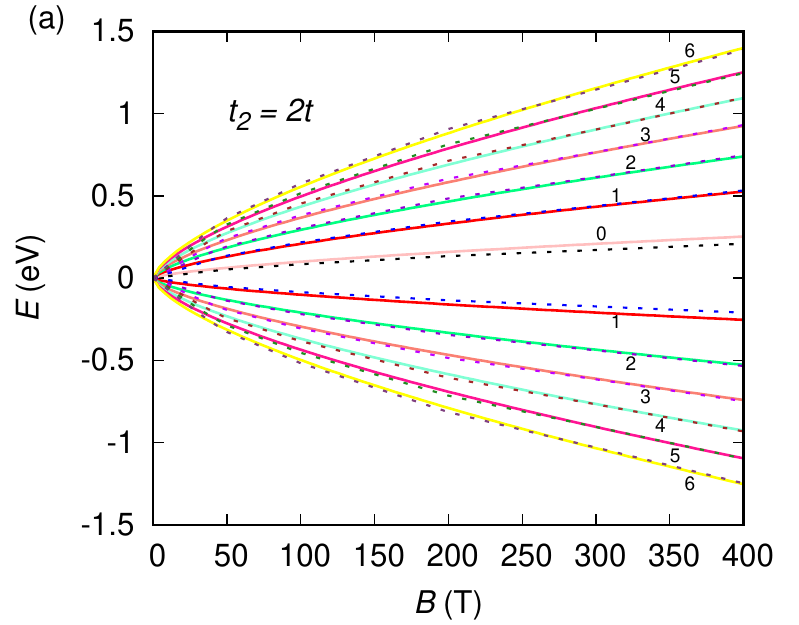}\label{fig:1a}} \vspace{-0.3 cm}
\subfloat{\includegraphics[width=0.4\textwidth]{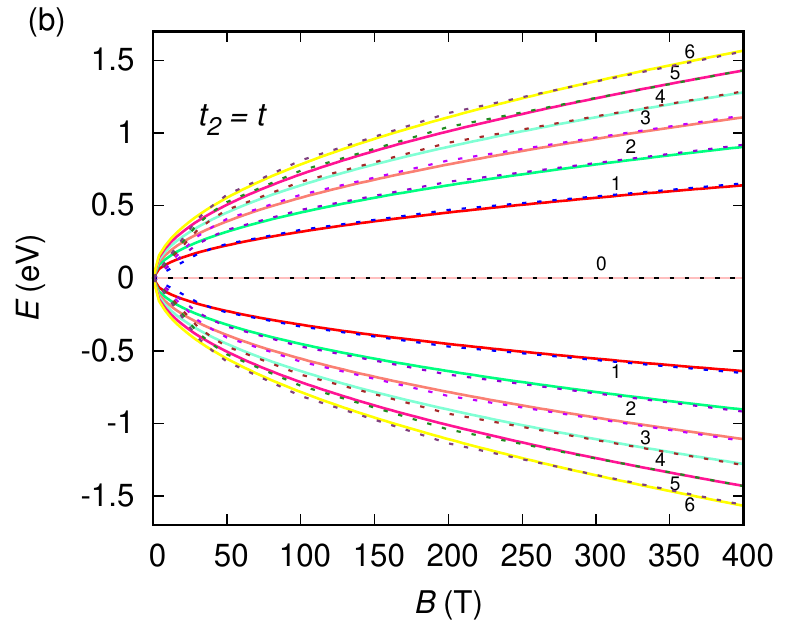}\label{fig:1b}}\\
\subfloat{\includegraphics[width=0.4\textwidth]{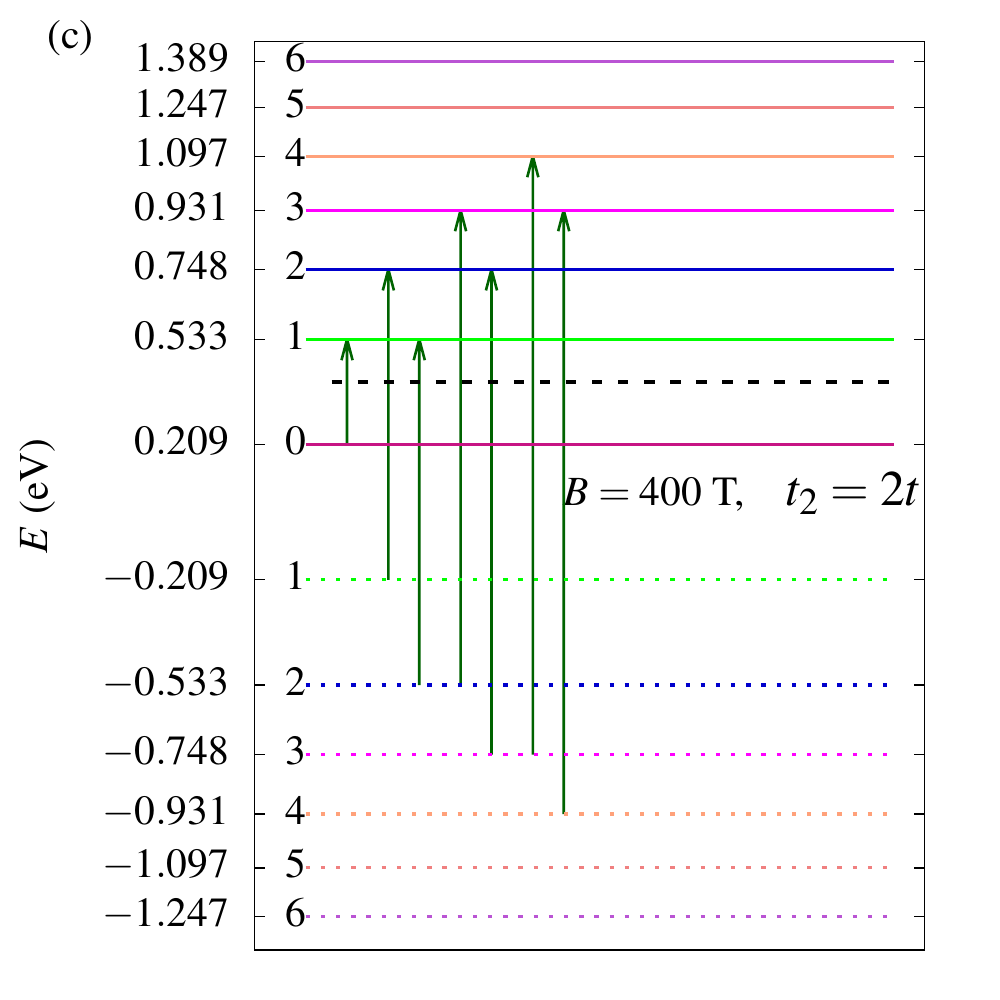}\label{fig:1c}} \vspace{-0.3 cm}
\subfloat{\includegraphics[width=0.4\textwidth]{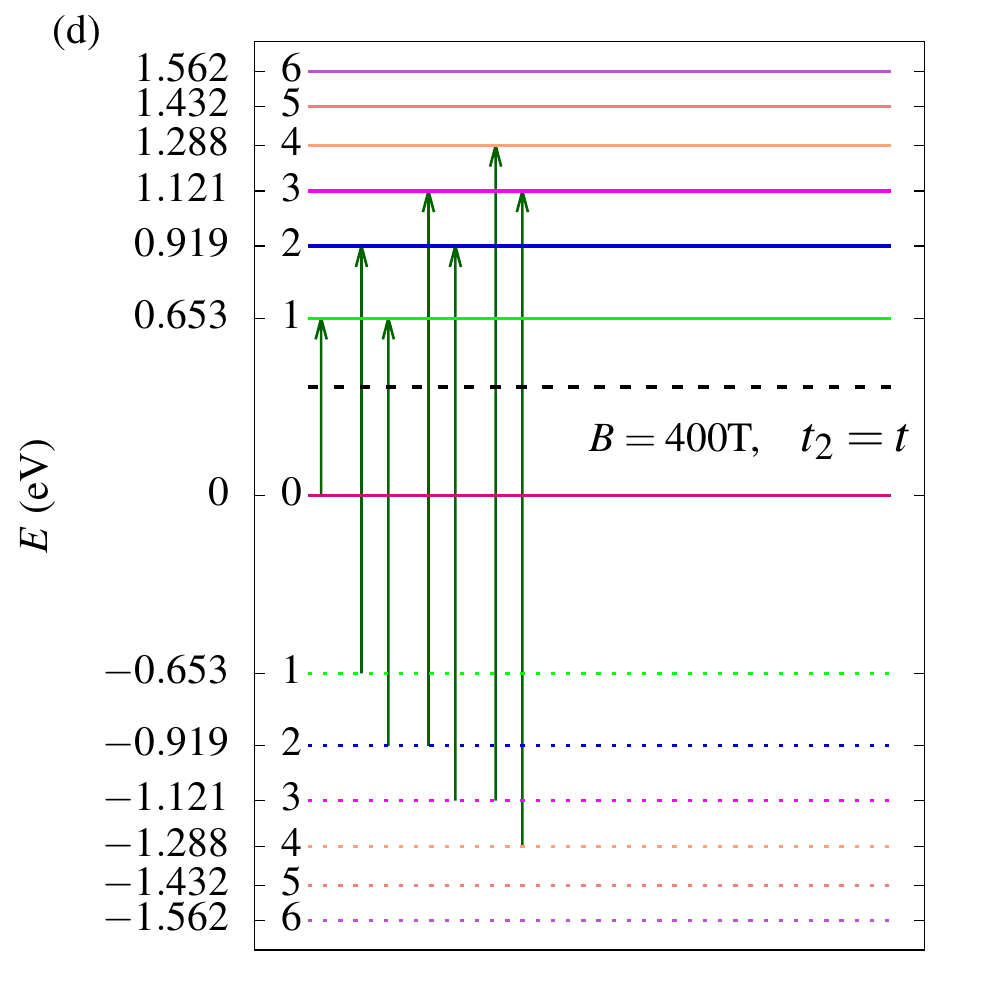}\label{fig:1d}}\\ 
\subfloat{\includegraphics[width=0.4\textwidth]{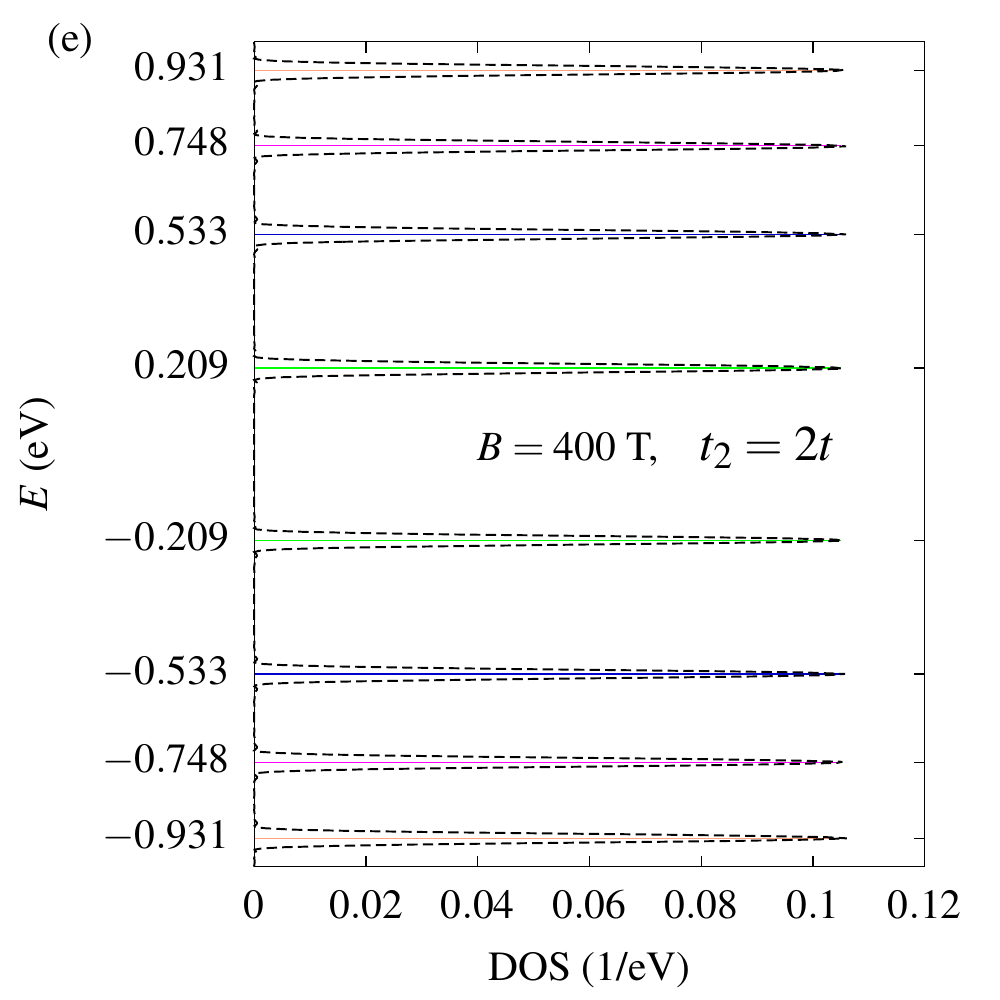}\label{fig:1e}} 
\subfloat{\includegraphics[width=0.4\textwidth]{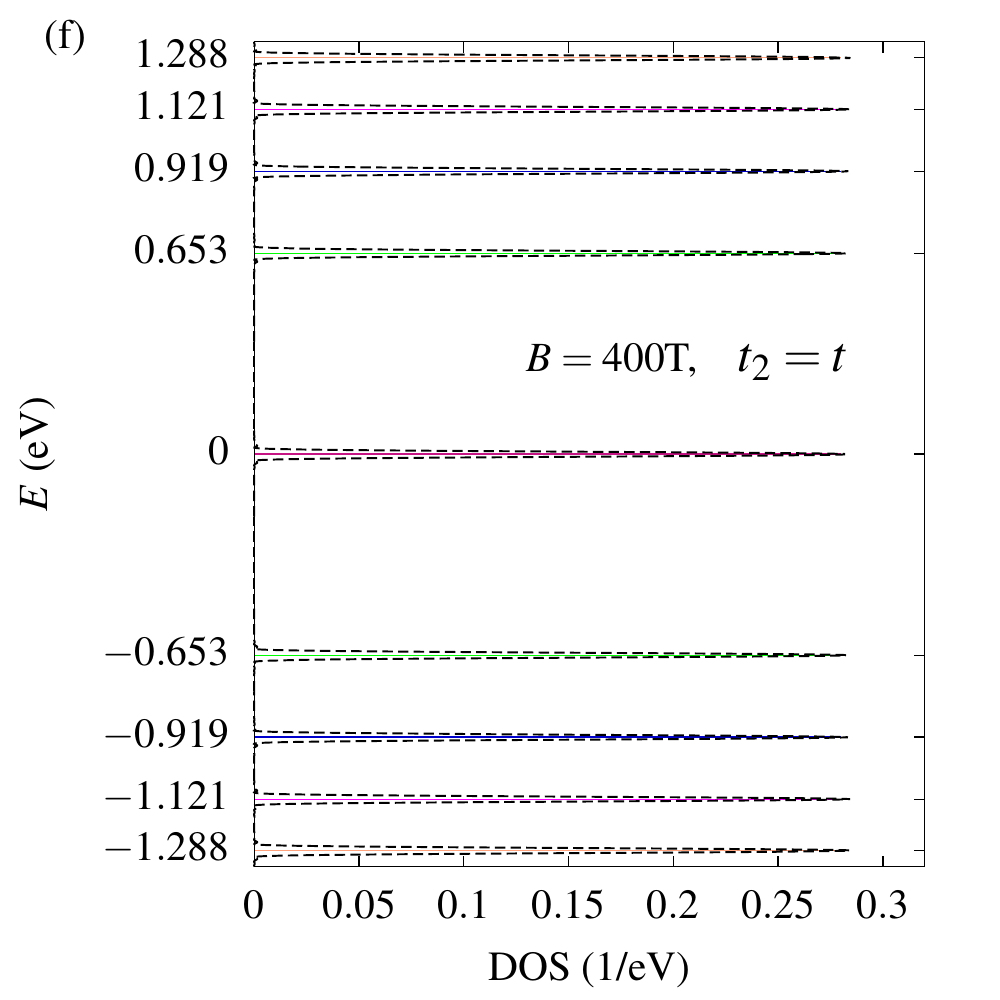}\label{fig:1f}}
\caption{(Color online) The two upper panels give the Landau level energies, $E$ (in units of eV) as a function of the magnetic field, $B$ (in units of Tesla) for various values of Landau level indices $n$ (labelled as 0, 1, 2 , 3, 4, $\dotsb$) for (a) $t_2=2t$ (semi-Dirac) and (b) $t_2=t$ (Dirac). The solid and the dotted lines are obtained from theoretical scaling ($E$ goes as $B^{2/3}$ for semi-Dirac case and $\sqrt{B}$ for Dirac case) and simulation respectively. In the two middle panels ((c) and (d)), a few allowed optical transitions are indicated by the vertical (dark-green) arrows and the chemical potential ($\mu=0.4$ eV) is shown by the horizontal black dashed line. The left and right panels correspond to $t_2=2t$ and $t_2=t$ at $B=400$T respectively. In the two lower panels energy levels, $E$ (in units of eV) versus density of states (DOS) (in units of 1/eV) are shown for (e) $t_2=2t$ (semi-Dirac) and (f) $t_2=t$ (Dirac) at $B=400$T.}
\label{fig:1}
\end{center}
\end{figure*}
\begin{figure*}[!ht!]
\begin{center}
\subfloat[]{\includegraphics[width=0.48\textwidth]{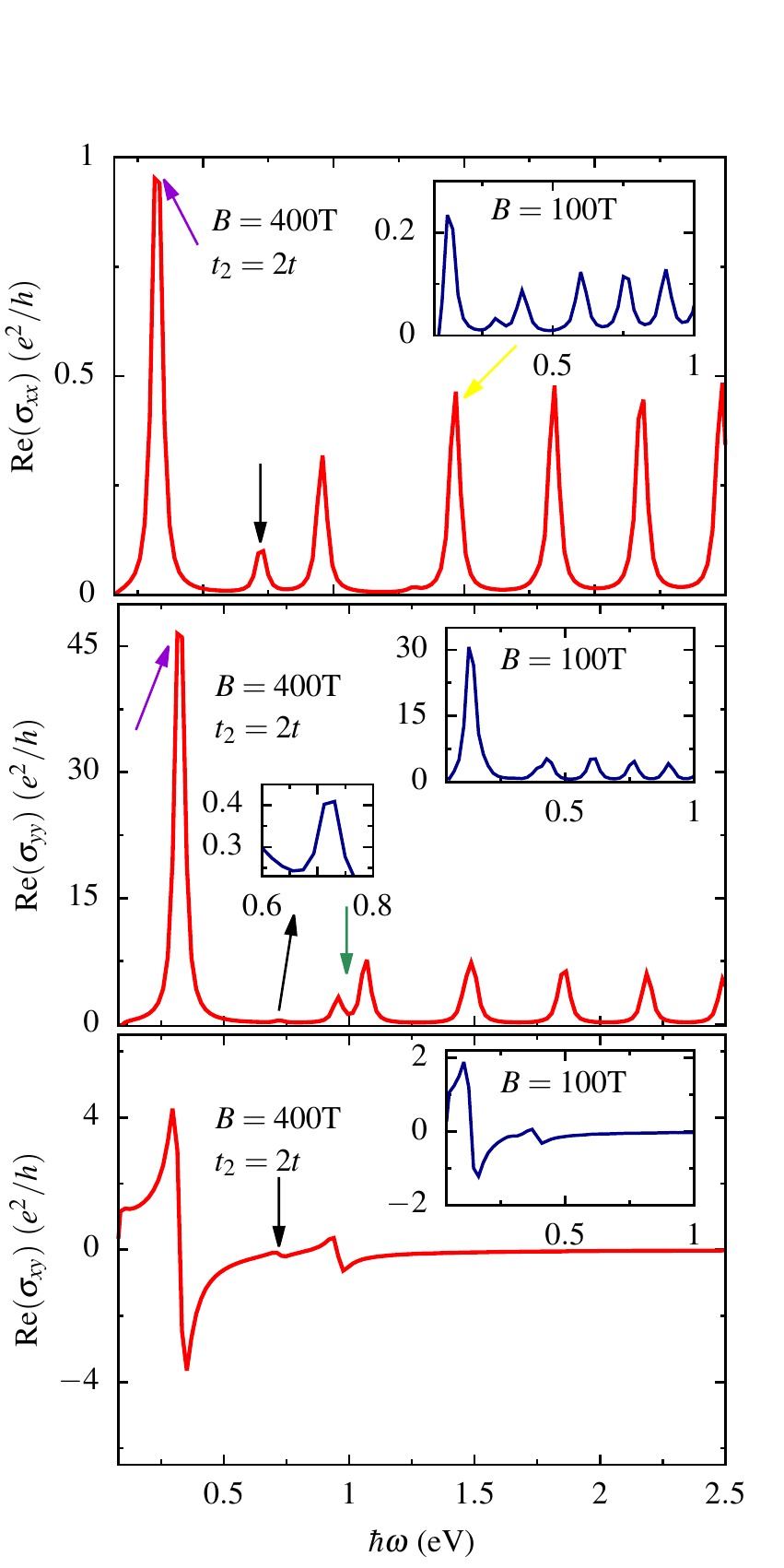}\label{fig:2a}} 
\subfloat[]{\includegraphics[width=0.48\textwidth]{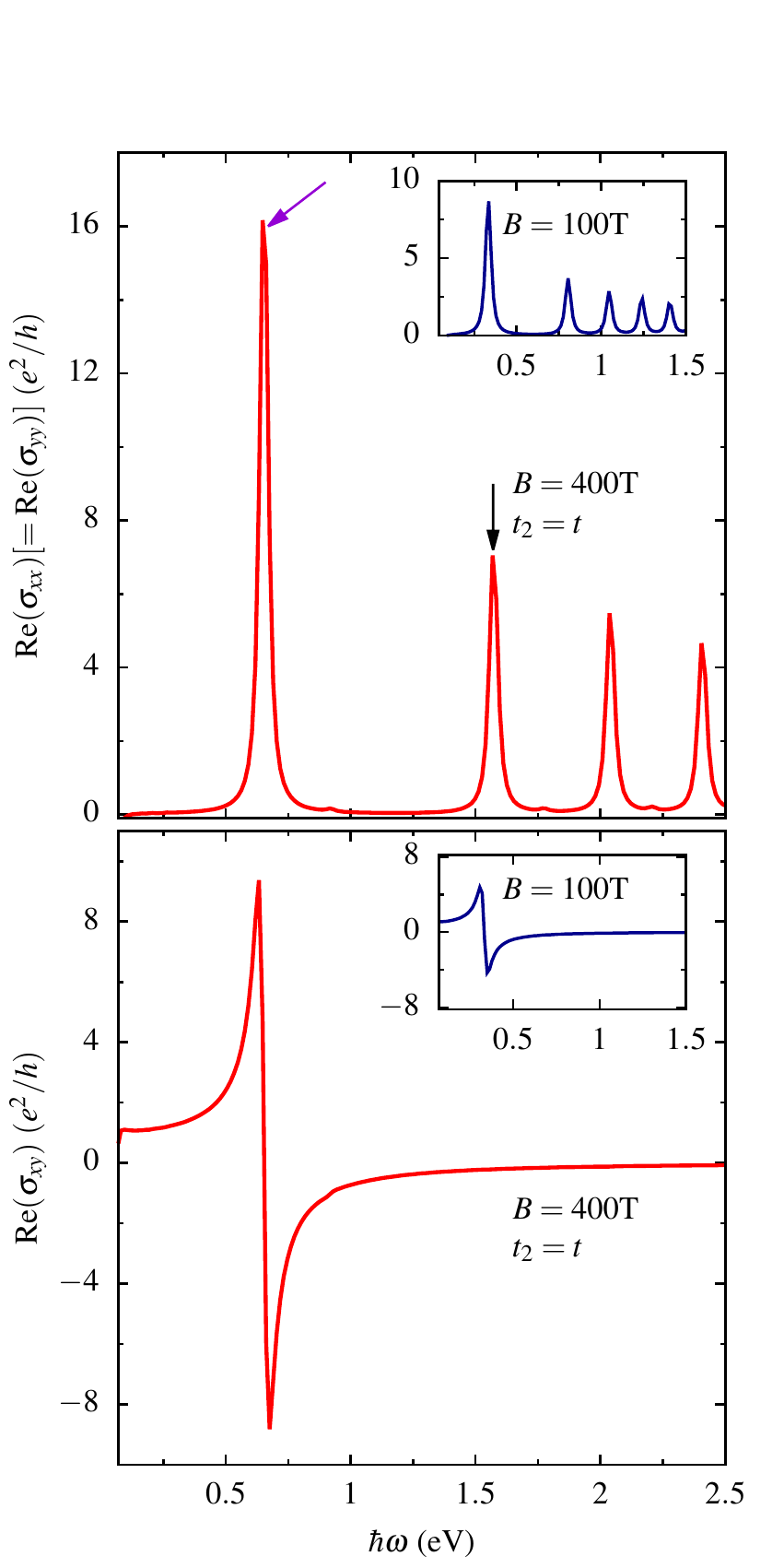}\label{fig:2b}} 
\caption{(Color online) The real parts of the longitudinal optical conductivities, $\sigma_{xx}$ and $\sigma_{yy}$ and the optical Hall conductivity, $\sigma_{xy}$ (in units of $e^2/h$) are shown as a function of photon energy, $\hbar\omega$ (in units of eV) for (a) $t_2=2t$ (semi-Dirac) and (b) $t_2=t$ (Dirac) at $B=400$T in the main frame (denoted by red curve). The inset plots show the same for more moderate values of magnetic field, say 100T (denoted by blue curve) for $t_2=2t$ and $t_2=t$. $\mu$ is set to be 0.4 eV.}
\label{fig:2}
\end{center}
\end{figure*}
\begin{figure*}[!ht!]
\begin{center}
\subfloat[]{\includegraphics[width=0.48\textwidth]{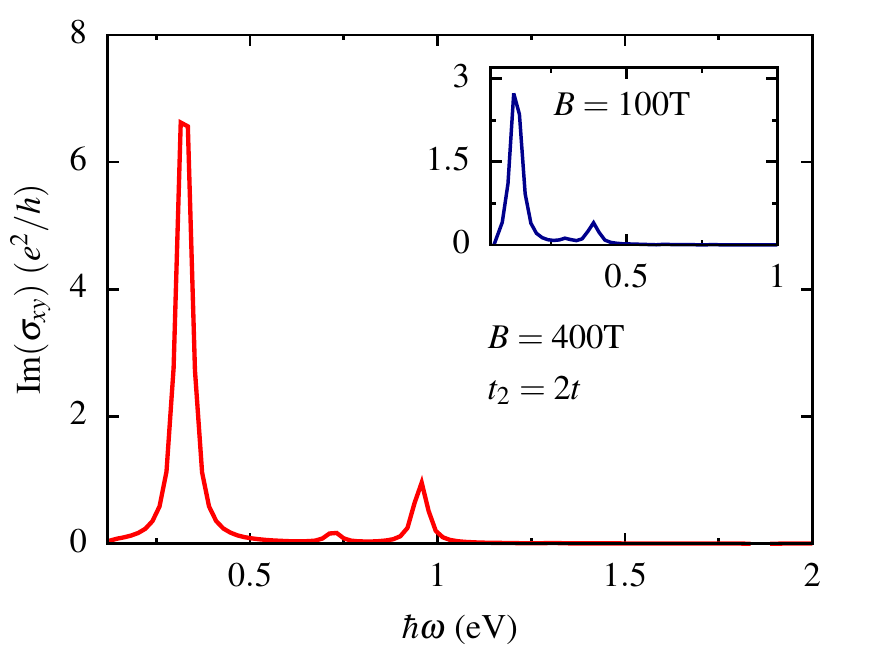}\label{fig:3a}} 
\subfloat[]{\includegraphics[width=0.48\textwidth]{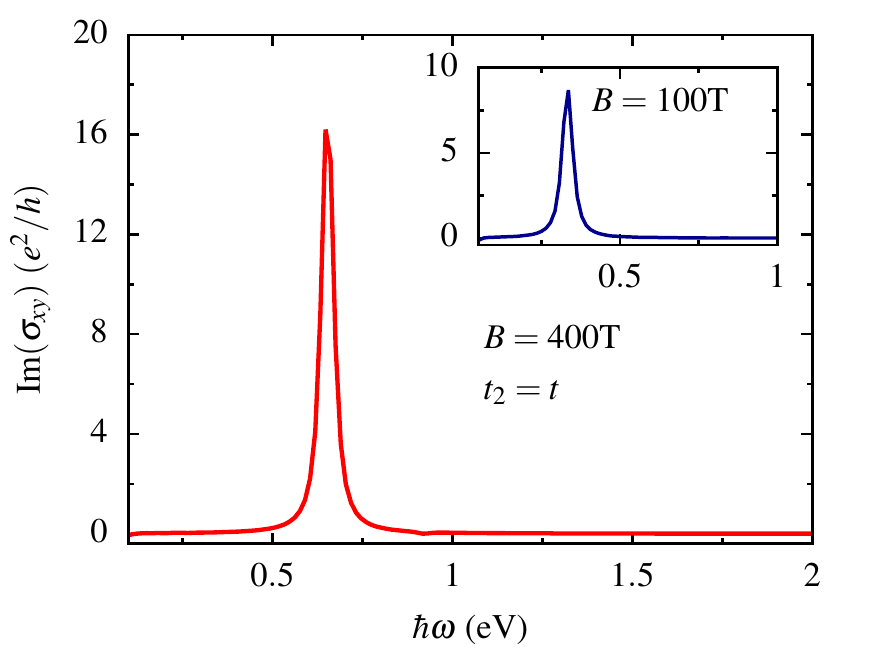}\label{fig:3b}}
\caption{(Color online) The imaginary parts of the optical Hall conductivity, $\sigma_{xy}$ (in units of $e^2/h$) are shown as a function of photon energy, $\hbar\omega$ (in units of eV) for (a) $t_2=2t$ (semi-Dirac) and (b) $t_2=t$ (Dirac) at $B=400$T in the main frame (shown by the red curve). The inset plots show the same for moderate values of magnetic field, (for example, $B=100$T) for $t_2=2t$ and $t_2=t$. $\mu$ is set to be 0.4 eV.}
\label{fig:3}
\end{center}
\end{figure*}
\subsection{Magnetotransport}\label{A}
To study the optical transport properties in the presence of a perpendicular magnetic field, we consider a semi-Dirac nanoribbon that consists of approximately $10^6$ number of atoms. In presence of a magnetic field, the off-diagonal terms in the $xy$ and $yx$ directions ($\sigma_{xy}$ and $\sigma_{yx}$) and the diagonal terms in $xx$ and $yy$ directions ($\sigma_{xx}$ and $\sigma_{yy}$) both contribute to the optical transport as seen from Eq.~(\ref{eq:12}).\par 
In the following, we wish to discuss the effect of magnetic field on a semi-Dirac nanoribbon. Results for a Dirac ribbon are included for comparison all the while. In Figs.~\ref{fig:1a} and \ref{fig:1b} we have shown the Landau levels, $E_n$ (both above and below the zero energy) for different Landau level indices, $n$ ($n$ = 0, 1, 2, 3, 4 $\dotsb$) as a function of the magnetic field, $B$ (in Tesla) for $t_2=2t$ (semi-Dirac) and $t_2=t$ (Dirac) cases. It is known that the Landau levels for a semi-Dirac system\cite{pickett} depend on the index, $n$ and the magnetic field, $B$ via $\big((|n+\frac{1}{2}|)B\big)^{2/3}$, while the corresponding dependence for a Dirac system \cite{sadowski2} are more well-known, namely, $(|n|B)^{1/2}$. Thus these analytic forms can be used to compare with the numerical values obtained by us. In the upper panel of Fig.~\ref{fig:1}, we show these analytic forms via solid lines, while the numerical results are demonstrated via dotted lines. It is seen that the agreement is fairly good in both the cases, which essentially becomes perfect for large values of $n$ for the semi-Dirac case (for the Dirac case, we have a fairly good agreement for all values of $n$).\par 
Let us look at the plots more closely. For the semi-Dirac case, the solid pink curve and the dashed black curve that correspond to the lowest Landau level ($n=0$), are slightly shifted from $E=0$ for all values of the magnetic field as seen from Fig.~\ref{fig:1a}. This is in contrast to the Dirac case, where the energy scales as $\sqrt{B}$ and the solid lines coincide with the dotted ones for positive as well as negative energy levels, with the $n=0$ Landau level occurring exactly at zero energy as shown in Fig.~\ref{fig:1b}. A particle-hole symmetry with respect to $E=0$ is preserved for the Dirac case. The particle-hole symmetry corresponding to the positive and negative energy levels is no longer observed for the semi-Dirac case ($t_2=2t$). In the middle panel of Fig.~\ref{fig:1} (Fig.~\ref{fig:1c} and Fig.~\ref{fig:1d}), we show possible optical transitions at a particular value of the magnetic field, namely, $B=400$T for two different systems at a fixed value of the chemical potential, $\mu=0.4$ eV. The solid lines denote the positive branches, whereas the negative branches are denoted by dotted lines. Apart from that, the arrows in the middle panel depict the transition from the occupied to the unoccupied levels through the absorption of a photon. The value of the chemical potential used in our computations is shown by a horizontal black dashed line which falls between the two consecutive Landau levels. The effects of varying the chemical potential will be discussed later.\par 
In Fig.~\ref{fig:2} we have shown the real parts of the longitudinal optical conductivities, Re($\sigma_{xx}$) and Re($\sigma_{yy}$) as well as the optical Hall conductivity, Re($\sigma_{xy}$) as a function of photon energy ($\hbar\omega$) for the semi-Dirac ($t_2=2t)$ and the Dirac ($t_2=t$) systems corresponding to a fixed chemical potential $\mu=0.4$ eV at $B=400$T (shown by the red curve) in the main frame. Plots with a more moderate value of the magnetic field (say, 100T) are shown in the inset of Fig.~\ref{fig:2} (shown by the blue curve). The real parts are related to the optical absorption of the nanoribbon and hence characterize the MO properties.\par The asymmetry in the semi-Dirac case mentioned above has a consequence on the transport properties presented below. Particularly, the peaks for the real as well as the imaginary parts of the MO conductivity are modified compared to the Dirac case. This is depicted in Fig.~\ref{fig:2} and Fig.~\ref{fig:3}. For the real parts of $\sigma_{xx}$ and $\sigma_{yy}$, a series of asymmetric absorption resonance peaks are observed for both the semi-Dirac ($t_2=2t$) and the Dirac ($t_2=t$) cases, which result from the optical transitions between different Landau levels. Since in optical transitions, the selection rules allow the value of $n$ to change only by 1, the transition from $n$ = 0 to $n$ = 1, indicated by the shortest green arrow in Figs.~\ref{fig:1c} and \ref{fig:1d} gives the first peak (denoted by dark-violet arrow) in Re($\sigma_{xx}$) for both $t_2=2t$ and $t_2=t$ as shown in the top panels of Fig.~\ref{fig:2} (Figs.~\ref{fig:2a} and \ref{fig:2b}). The only difference that can be seen is that the peak has shifted slightly to lower energy with reduced intensity for the semi-Dirac case ($t_2=2t$).\par Next, we observe that the two arrows, that is, from $n$ = 1 (negative energy) to $n$ = 2 (positive energy) and from $n$ = 2 (negative energy) to $n$ = 1 (positive energy) contribute to the formation of the second peak (denoted by a black arrow). For the Dirac case ($t_2=t$), these two arrows have exactly the same length and consequently, there is only one peak in the conductivity spectrum. In contrast to the Dirac case, these two arrows have slightly different lengths for the semi-Dirac case ($t_2=2t$), since the symmetry between the positive and the negative spectra ceases to exist. Still we have observed a single peak (denoted by black arrow) in Re($\sigma_{xx}$) because the energy difference between a transition from $n$ = 1 (negative side) to $n$ = 2 (positive side) and that from $n$ = 2 (negative side) to $n$ = 1 (positive side) is negligibly small. The rest of the peaks (third, fourth, and so on) are similar to the second peak; they come from a pair of transitions from $-n$ to $n$ + 1 and $-(n$ + 1) to $n$ as shown in Fig.~\ref{fig:1}. For example, the fourth peak (denoted by yellow arrow) observed in upper panel of Fig.~\ref{fig:2a} is due to the combined transitions from $n$ = 4 (negative) to $n$ = 3 (positive) and $n$ = 3 (negative) to $n$ = 4 (positive). For the Dirac case ($t_2=t$), the real part of $\sigma_{yy}$ gives the same result as that of $\sigma_{xx}$ due to the isotropic nature of the system. Nevertheless, we show that both $\sigma_{xx}$ and $\sigma_{yy}$ in the same plot (as shown in the upper panel of Fig.~\ref{fig:2b}). Whereas for the semi-Dirac case ($t_2=2t$), we have $\sigma_{xx} \neq \sigma_{yy}$ owing to the anisotropic band dispersion along the $k_x$ and the $k_y$ directions (see the middle panel of Fig.~\ref{fig:2a}). In this case, the intensity of the absorption peak for Re($\sigma_{yy}$) is much larger (roughly one order of magnitude) than those of Re($\sigma_{xx}$). Also, the height of the second peak (indicated by black arrow) in the Re($\sigma_{yy}$) is too smaller compared to other peaks as shown in the inset plot,  whereas the third one (denoted by green arrow) splits due to the energy difference as mentioned earlier. The peak positions and the intensities of the transport phenomena are functions of both the velocities of the electrons in the Landau levels and electron filling. For a given Landau level, the carriers in the semi-Dirac case have lesser velocity. This causes a lower peak height than the Dirac case. As mentioned earlier, the electron density plays a role as well in shaping the peaks observed in the real parts of $\sigma_{xx}$ and $\sigma_{yy}$ for both the semi-Dirac and Dirac cases. This can be seen via the density of states (DOS) plotted in the lowest panels of Fig.~\ref{fig:1}, namely Figs.~\ref{fig:1e} and \ref{fig:1f}. The magnitude of the DOS plotted along the $x$-axis corresponding to the semi-Dirac case is at least smaller by a factor of two than those for the Dirac case.\par
\begin{figure*}[!ht!]
\begin{center}
\subfloat[]{\includegraphics[width=0.48\textwidth]{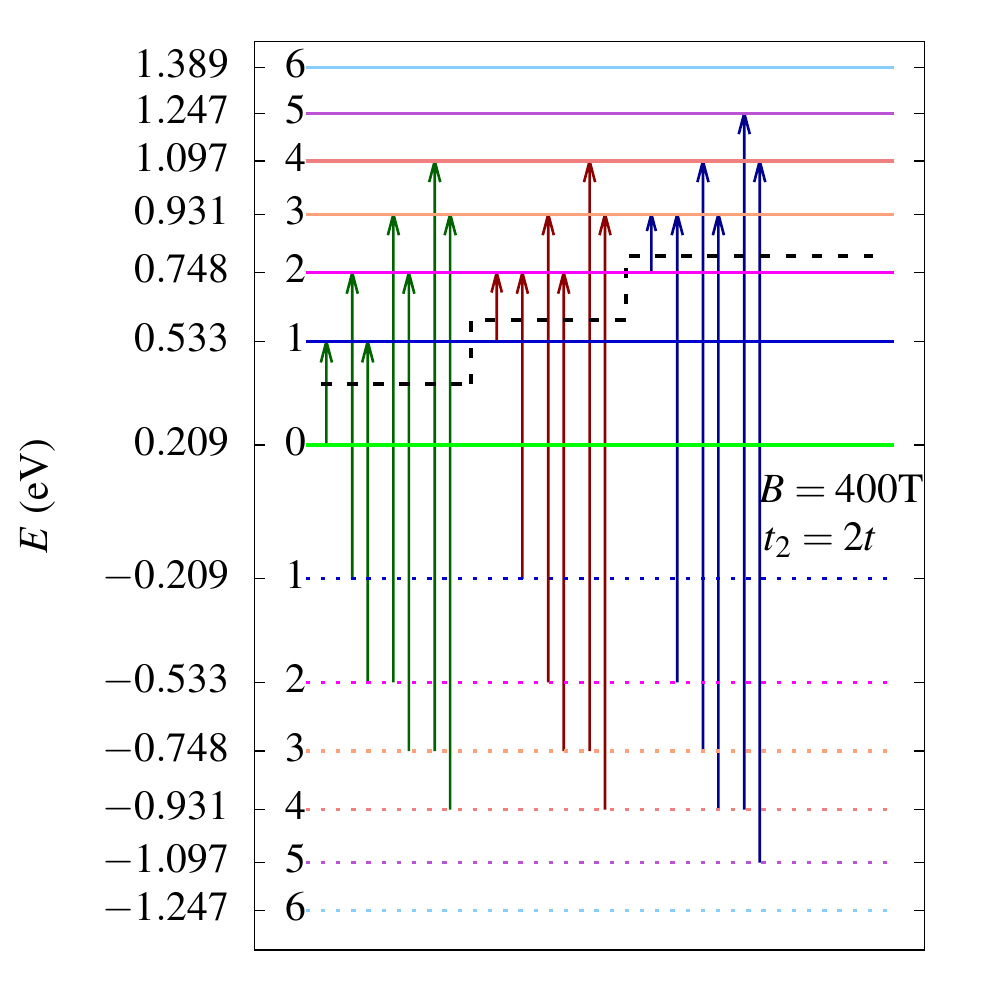}\label{fig:4a}} 
\subfloat[]{\includegraphics[width=0.48\textwidth]{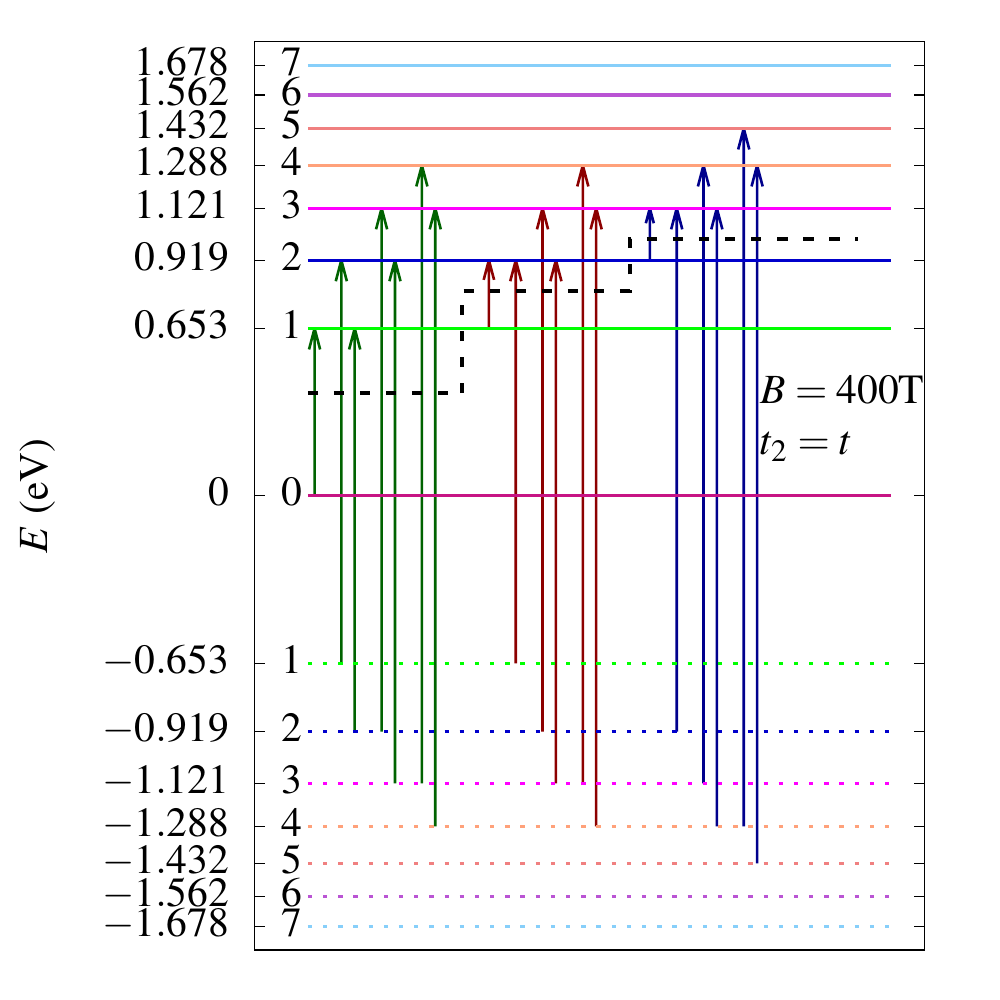}\label{fig:4b}}\\
\subfloat[]{\includegraphics[width=0.48\textwidth]{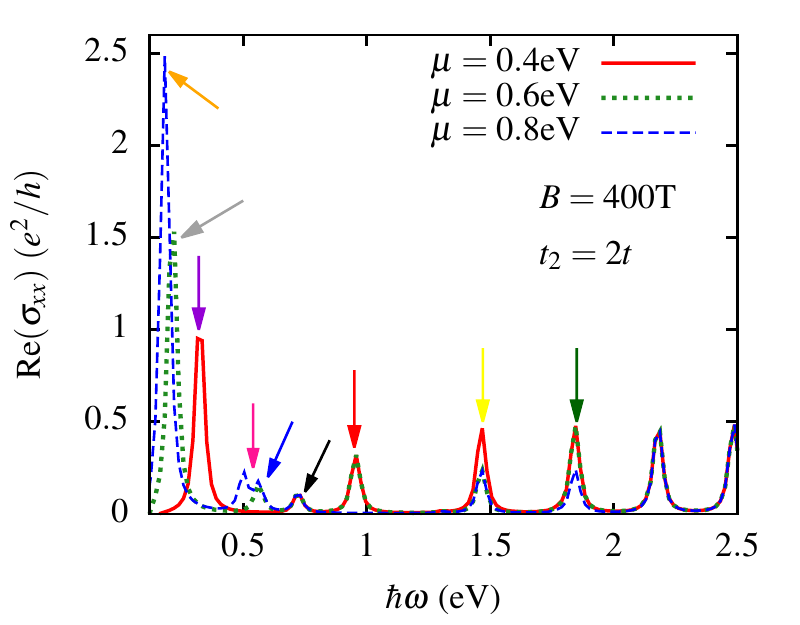}\label{fig:4c}} 
\subfloat[]{\includegraphics[width=0.48\textwidth]{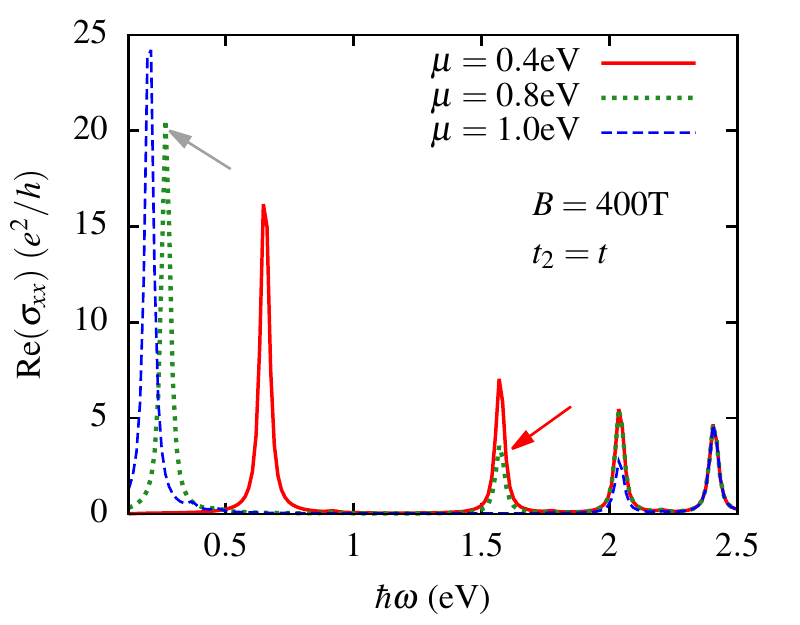}\label{fig:4d}}  
\caption{(Color online) In the upper two panels, a few allowed optical transitions are shown indicated by arrows for various values of the chemical potential for (a) $t_2=2t$ (semi-Dirac) and (b) $t_2=t$ (Dirac) at $B=400$T. Those $\mu$ values are marked with the horizontal black dashed lines. In the lower two panels, the real parts of the longitudinal optical conductivity, $\sigma_{xx}$ in units of $e^2/h$ are shown as a function of photon energy, $\hbar\omega$ in units of eV for various values of chemical potential $\mu$ for (c) $t_2=2t$ (semi-Dirac) and (d) $t_2=t$ (Dirac) at $B=400$T.}
\label{fig:4}
\end{center}
\end{figure*}
So far, we have discussed the MO conductivity considering the diagonal term, namely $\sigma_{xx}$ and $\sigma_{yy}$. It is also of interest to see the effects of the off-diagonal component, namely $\sigma_{xy}$ which we shall discuss here. In the bottom panel of Fig.~\ref{fig:2} we have plotted the real part of the optical Hall conductivity, Re($\sigma_{xy}$) as a function of the photon energy for both the semi-Dirac ($t_2=2t$) and the Dirac ($t_2=t$) cases at $B=400$T (shown by the red curve) as shown in the main frame. The main features of the real part of the Hall response are its antisymmetric behavior about its zero value and the presence of a single peak on either side of zero intensity. The peak in the spectrum results from a single transition (that is, $n=0$ to $n=1$) that contributes to the Hall conductivity $\sigma_{xy}$. In the lower panel of Fig.~\ref{fig:2a} for the semi-Dirac case ($t_2=2t$), the first peak (in the positive direction) with maximum intensity occurs at 0.29 eV and the first peak (in the negative direction) with a minimum intensity occurs at 0.35 eV, both of which are shifted to lower energies as compared to the Dirac case. Also, two pairs of positive and negative peaks are observed in the vicinity of 0.72 eV (denoted by black arrow) and 0.95 eV for the semi-Dirac case respectively. For the Dirac case ($t_2=t$), Re($\sigma_{xy}$) first shows a positive peak with maximum intensity occurring at 0.63 eV and hence a negative peak with minimum intensity occurring at 0.67 eV as shown in the lower panel of Fig.~\ref{fig:2b}. The similarity that exists between the semi-Dirac and the Dirac cases is that the Hall conductivity remains flat for low energies and tends to vanish for higher energies in both the cases.\par
Next, we have shown the imaginary part of the optical Hall conductivity, namely Im($\sigma_{xy}$) as a function of the photon energy, $\hbar\omega$ in Fig.~\ref{fig:3}. For the semi-Dirac case ($t_2=2t$), the imaginary part of $\sigma_{xy}$ only shows positive peaks (see Fig.~\ref{fig:3a}) exactly where the real part of the Hall conductivity, that is, Re($\sigma_{xy}$) shows the absorption peaks as seen from lower panel of Fig.~\ref{fig:2a}. In the case of the Dirac system ($t_2=t)$, Im($\sigma_{xy}$) shows a sharp single positive peak at the same energy that corresponds to the peak of the Re($\sigma_{xy}$) as shown in Fig.~\ref{fig:2b}. The peak positions for the real and the imaginary parts of the Hall conductivity, $\sigma_{xy}$ have a correspondence with the absorption peaks of the longitudinal conductivities for both the semi-Dirac and the Dirac cases as shown in Figs.~\ref{fig:2} and \ref{fig:3}. For example, the second peak of the real and imaginary part of the Hall conductivity, Re($\sigma_{xy}$) and Im($\sigma_{xy}$) occurs at an energy very close to 0.72 eV that correspond to the second peak (denoted by black arrow) for the Re($\sigma_{xx}$) for the semi-Dirac case as shown in Fig.~\ref{fig:2a}. This is expected as the absorption and the transmission both correspond to the transition of the carriers from one Landau level to another.\par
In the following, we observe the effects of a few key quantities that characterize the MO phenomena and also aid in distinguishing the semi-Dirac and the Dirac cases. Specifically, we explore effects of the electron filling via tuning the chemical potential, $\mu$ and the polarization of the incident light. Further, we study another quantity that characterizes the MO property, namely the Faraday rotation angle. 
\subsection{Electron Filling}\label{B}
In this section, we shall see how the electron concentration affects the MO conductivities. The electron densities can be controlled by varying the chemical potential, $\mu$ which can further be tuned using external means, for example, a gate voltage. The values of $\mu$ considered here are different for the Dirac and the semi-Dirac cases. However, care is taken so that the corresponding values lie between same pairs of the Landau level in both cases. As $\mu$ is varied, it moves through the successive Landau levels. In Figs.~\ref{fig:4a} and \ref{fig:4b}, we have shown the allowed transitions for different values of the chemical potentials by arrows of different colors at $B= 400$T for the semi-Dirac ($t_2 = 2t$) and the Dirac ($t_2 = t$) cases respectively. According to the transition rules, when $\mu$ falls between the $n^{\text{th}}$ and $(n+1)^{\text{th}}$ Landau levels, the transitions from values lower $n$ are blocked. So only the transitions passing through the $\mu$ value are allowed. For example, when $\mu$ lies between the Landau levels $n=1$ and $n=2$, the transition shown by the shortest dark-red arrow in Fig.~\ref{fig:4b} is allowed, whereas the transitions originating from Landau levels lower than $n=1$ is blocked. For example, transition from $n=0$ to $n=1$ is forbidden. These values of the chemical potential are demonstrated in the upper panel of the Fig.~\ref{fig:4} via black dashed lines. In our case, we have chosen three different values of $\mu=0.4$ eV, $0.6$ eV and $0.8$ eV, which fall between the zeroth and the first (where the transitions are shown by dark-green arrow), the first and the second (where the transitions are shown by dark-red arrow), and the second and the third (where the transitions are shown by dark-blue arrow) Landau levels respectively. For example, in Fig.~\ref{fig:4c}, we have shown the Re($\sigma_{xx}$) for the three different values of $\mu$ =0.4 eV (falls between the zeroth and the first Landau level), 0.6 eV (falls between the first and the second Landau level) and 0.8 eV (falls between the second and the third Landau level) for the semi-Dirac case ($t_2 = 2t$). When $\mu$ = 0.4 eV, the transition from $n$ = 0 to $n$ = 1 (shortest dark-green arrow in Fig.~\ref{fig:4a}) yields the first peak (denoted by dark-violet arrow) at lower energies. As we increase $\mu$ to 0.6 eV (shown by the green dotted curve), only the first (denoted by grey arrow) and the second (denoted by blue arrow) peaks shift to lower energies, which occur at 0.22 eV and 0.56 eV. Whereas the third (denoted by black arrow) and the fourth (denoted by red arrow) peaks coincide with the second and the third peaks that correspond to those for $\mu=0.4$ eV (solid red curve). When $\mu$ is increased further, that is, to 0.8 eV, though the first peak (dashed blue curve denoted by orange arrow) shifts to lower frequencies, the second peak (denoted by pink arrow) splits, in addition to getting shifted towards lower frequencies. Further, the third peak at an energy value 0.72 eV remains intact, whereas the fourth one (at an energy value 0.96 eV) vanishes.\par 
To compare with the Dirac case\cite{gusy1} ($t_2=t$), we show that Re($\sigma_{xx}$) as a function of the photon energy, $\hbar\omega$ for increasing values of the chemical potential $\mu$ = 0.4 eV (falls between the zeroth and the first Landau level), 0.8 eV (falls between the first and the second Landau level) and 1 eV (falls between the second and the third Landau level) in Fig.~\ref{fig:4d}. Here, the effects of $\mu$ shown by the solid red curve for $\mu$ = 0.4 eV (which falls between the zeroth and the first Landau level) and the corresponding results are already discussed in the previous section. If we increase $\mu$ from 0.4 eV to 0.8 eV, the first peak (dotted green curve denoted by grey arrow) that results from transition between the $n$ = 1 to $n$ = 2 levels (shown by the shortest dark-red arrow in Fig.~\ref{fig:4a}) shifts to lower frequencies. Compared to the first peak, the second one (denoted by red arrow) does not shift, however the intensity becomes half for the peak at an energy 1.57 eV. The reason is obvious since the transition from $n=-1$ to $n$ = 2 is allowed, whereas $n=-2$ to $n$ = 1 ($n$ = 1 level falls below the Landau level) is Pauli blocked. When $\mu$ is further increased to 1 eV (which falls between the second and the third Landau levels), the peak disappears since both the transitions ($n=-1$ to $n$ = 2 and $n=-2$ to $n$ = 1) are forbidden. Further, there is a reduction in the height of the peak at about 1.57 eV. 
\begin{figure*}[!ht!]
\begin{center}
\subfloat[]{\includegraphics[width=0.48\textwidth]{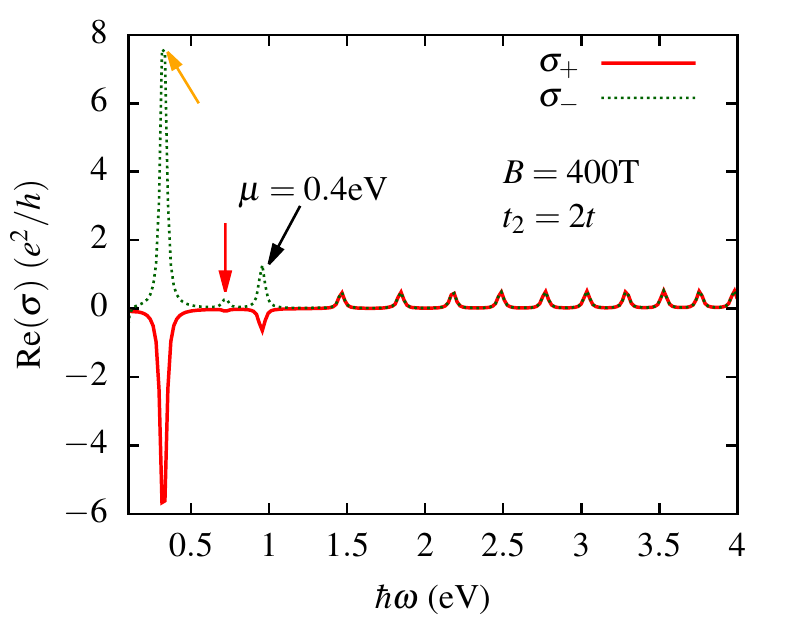}\label{fig:5a}} 
\subfloat[]{\includegraphics[width=0.48\textwidth]{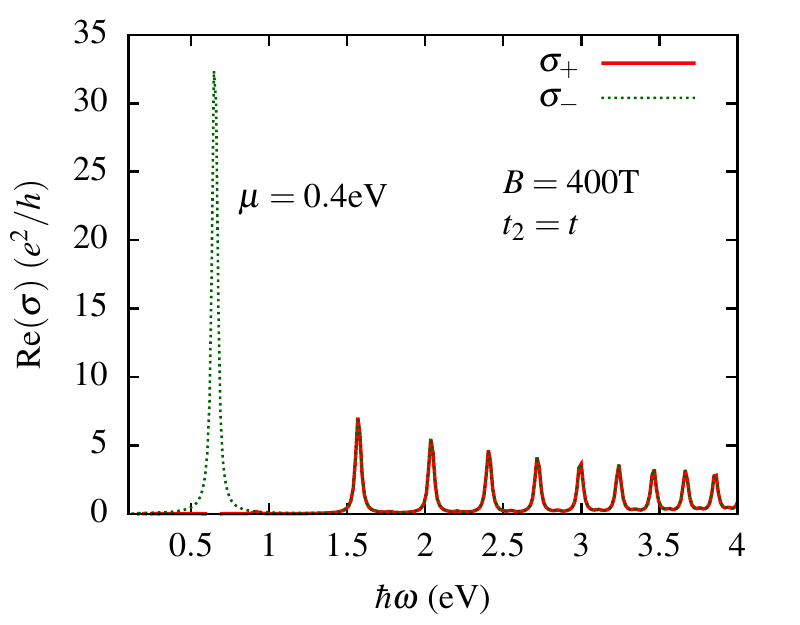}\label{fig:5b}}\\
\subfloat[]{\includegraphics[width=0.48\textwidth]{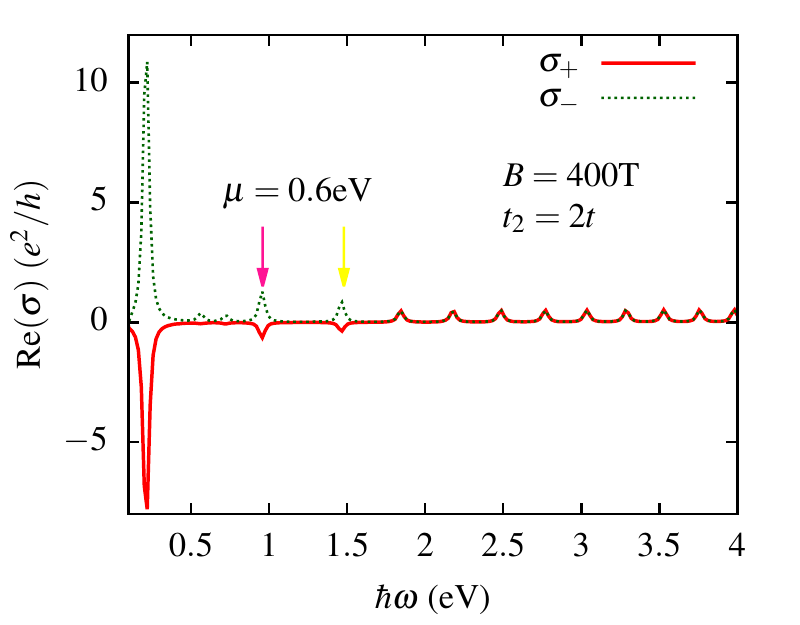}\label{fig:5c}} 
\subfloat[]{\includegraphics[width=0.48\textwidth]{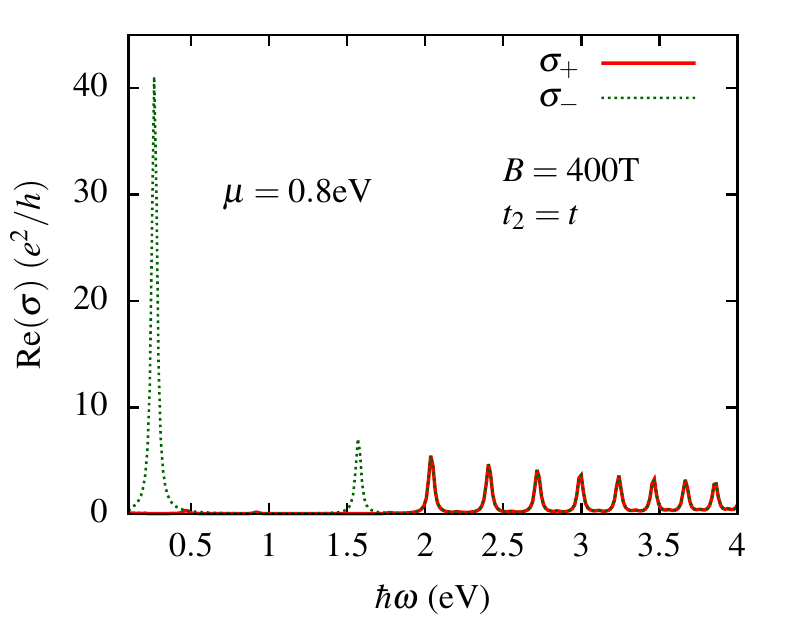}\label{fig:5d}}
\caption{(Color online) In the upper two panels, the real parts of the right-handed polarized optical conductivity $\sigma_{+}$ and that of the left-handed polarized one $\sigma_{-}$ (in units of $e^2/h$) are shown as a function of photon energy, $\hbar\omega$ (in units of eV) for a fixed value of chemical potential $\mu=0.4$eV for (a) $t_2=2t$ (semi-Dirac) and (b) $t_2=t$ (Dirac) at $B=400$T. The lower two panels ((c) and (d)) depict the same scenario by varying the chemical potential which falls between the $n=1$ and $n=2$ Landau levels.}
\label{fig:5}
\end{center}
\end{figure*}

\begin{figure*}[!ht!]
\begin{center}
\subfloat[]{\includegraphics[width=0.49\textwidth]{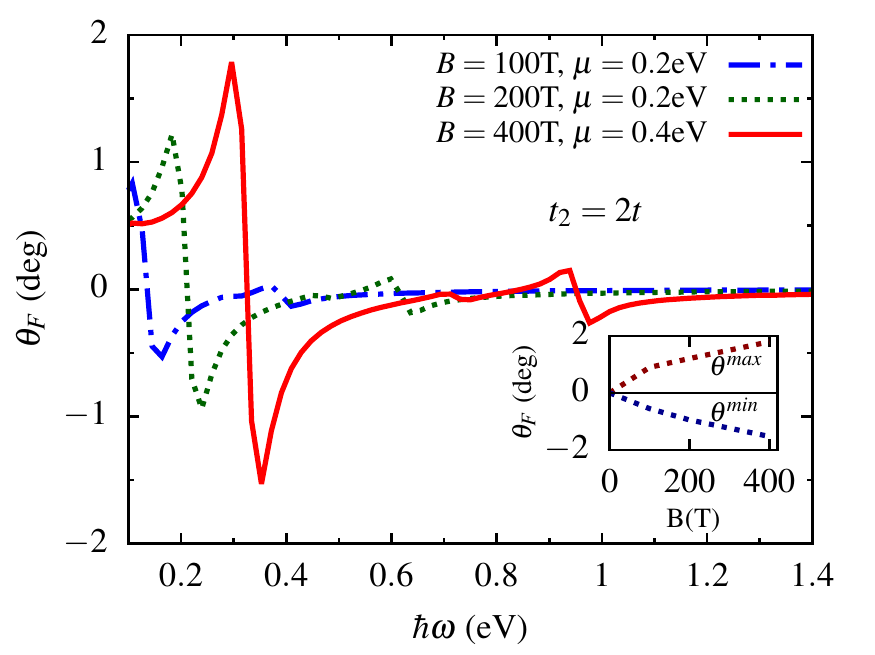}\label{fig:6a}} 
\subfloat[]{\includegraphics[width=0.49\textwidth]{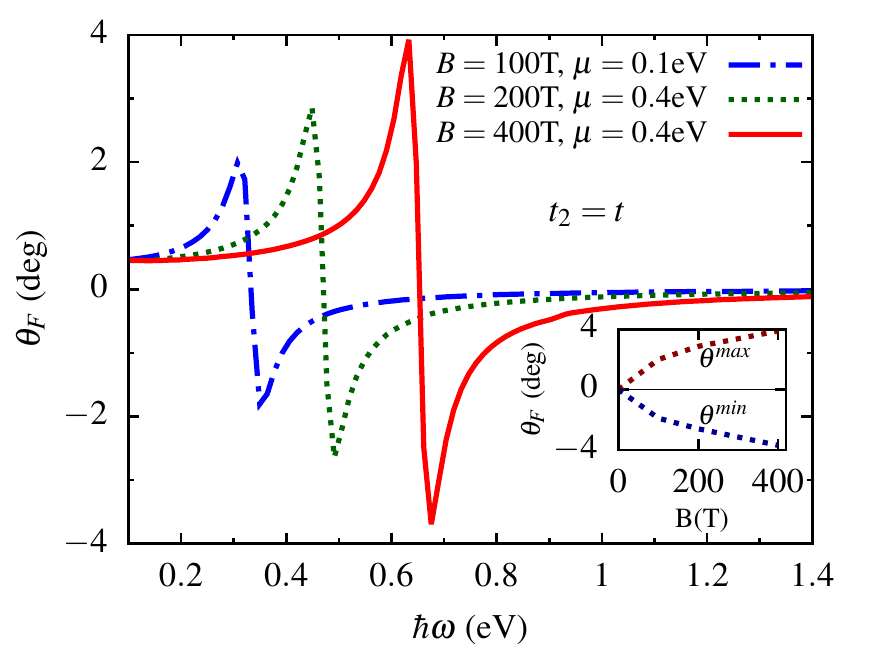}\label{fig:6b}}
\caption{(Color online) The Faraday rotation angle, $\theta_{F}$ (in units of deg) is plotted as a function of photon energy, $\hbar\omega$ (in units of eV) for different values of magnetic field and chemical potential for (a) $t_2=2t$ (semi-Dirac) and (b) $t_2=t$ (Dirac). The insets show the maximum ($\theta^{max}$) and the minimum ($\theta^{min}$) Faraday angle, $\theta_{F}$ (in units of deg) versus magnetic field, $B$ (in units of Tesla).}.
\label{fig:6}
\end{center}
\end{figure*}
\subsection{Circular Polarization}\label{C}
Here, we probe the effects of changing the polarization of the incident light as usually done in experiments. Instead of a linearly polarized light, we can take a circularly polarized one, whose effects can be simply incorporated by superposing the quantities obtained in our MO transport studies. For example, we can define $\sigma_{\pm}$=$\sigma_{xx}\pm i\sigma_{xy}$, where $\sigma_+$ denotes light with right-handed polarization and $\sigma_{-}$ denotes that with left-handed polarization. The real (absorptive) part of $\sigma_{\pm}$ can be written as,
\begin{equation}
\text{Re}(\sigma_{\pm}(\omega)) = \text{Re}(\sigma_{xx}(\omega))\mp \text{Im}(\sigma_{xy}(\omega)),
\label{eq:13}
\end{equation}
where the upper(lower) sign corresponds to conductivities with the right(left) circular polarization. In Fig.~\ref{fig:5}, we show the Re($\sigma$) as a function of the photon energy, $\hbar\omega$ for $t_2=2t$ and $t_2=t$ at $B=400$T. The dotted green curve is for $\sigma_{-}$ and the solid red curve is for $\sigma_{+}$. In the upper panels, $\mu$  is taken to be 0.4 eV which lies between the zeroth and the first Landau levels. For the semi-Dirac case ($t_2=2t$), the lowest frequency peaks in Re($\sigma_{-}$) differ significantly than in Re($\sigma_{+}$). It can be noted from Fig.~\ref{fig:5a} that only the peaks with the first (denoted by orange arrow), second (denoted by red arrow) and third (denoted by black arrow) lowest frequencies become positive for $\sigma_{-}$, whereas they are negative for $\sigma_{+}$. Moreover, the magnitude of those three peaks is lesser for $\sigma_{+}$ than those for $\sigma_{-}$. The rest of the peaks have almost the same magnitude. However, when $t_2=t$, Re($\sigma_{-}$) has only one peak at a low frequency which is absent in Re($\sigma_{+}$) due to the cancellation between the longitudinal and the transverse Hall conductivities with each other\cite{ashby,nicol,sun}. For the semi-Dirac case ($t_2=2t$), the anisotropy of the Dirac cone in the band structure leads to non-cancellation between the longitudinal and the transverse Hall conductivities. This low-frequency peak in Re($\sigma_{-}$) is followed by a series of interband peaks that are identical to those in Re($\sigma_{+}$) as shown in Fig.~\ref{fig:5b}. When the chemical potential is increased such that it lies between the first and the second Landau levels, the fourth (denoted by pink arrow) and fifth (denoted by yellow arrow) peak for $\sigma_{+}$ becomes negative with increased intensity for $t_2=2t$ as shown in Fig.~\ref{fig:5c}. For the Dirac case ($t_2=t$), the first peak at an energy value 1.57 eV in $\sigma_{+}$ vanishes as seen from Fig.~\ref{fig:5d}. Thus our studies imply that the polarization of the incident light brings in significant changes to the MO transport that can be easily probed in experiments. Other polarizations, such as elliptical polarization, etc. may have similar observable changes. 
\subsection{Faraday Rotation}\label{D}
Finally, we shall discuss Faraday rotation in a semi-Dirac system and compare it to the Dirac case. It is interesting to mention that the real part of transverse MO conductivity (that is, $\sigma_{xy}$) calculated here can be directly used in Faraday rotation experiments \cite{marel}. The Faraday-rotation angle $\theta_{F}$, which is proportional to $\sigma_{xy}$ can be written as,
\begin{equation}
\theta_{F}=\frac{2\pi}{c}\text{Re}(\sigma_{xy}(\omega)),
\end{equation}
where $c$ is the velocity of light in vacuum and Re($\sigma_{xy}$) is the real part of MO Hall conductivity. Figs.~\ref{fig:6a} and \ref{fig:6b} show the Faraday angle, $\theta_{F}$ as a function of the photon energy, $\hbar\omega$ in the terahertz range for different values of the magnetic field at a particular value of the chemical potential which falls between the zeroth and the first Landau level for the semi-Dirac ($t_2=2t$) and the Dirac ($t_2=t$) cases respectively. The corresponding values of $\mu$ at given values of $B$ are quoted in the plots. The spectra show an edge-like structure with a positive rotation at low energies and a negative rotation at higher energies for both the semi-Dirac and the Dirac cases. For the semi-Dirac case, the maximum peak of the Faraday rotation shifts towards lower energies, and the value of the maximum Faraday rotation is 1.8$^{\circ}$ ($\approx$ 0.03 rad) at $B=400$T which is smaller compared to the Dirac case as shown in Fig.~\ref{fig:6a}. It is to be noted that a discernible Faraday rotation is also observed at 0.93 eV with a maximum value of 0.17$^{\circ}$ ($\approx$ 0.003 rad) at $B=400$T which is absent for the Dirac case. For smaller values of magnetic field (say, for $200$T and $100$T), the plot shows a qualitatively similar behavior. For the Dirac case, the maximum Faraday rotation is 3.9$^{\circ}$ ($\approx$ 0.06 rad) at $B=400$T, which is larger than the other two magnetic fields, namely, $B=200$T (shown by green dotted curve) and 100T (shown by red dashed curve) as shown in Fig.~\ref{fig:6b}. For both the cases, the spectra depend on the value of the magnetic field used. To further endorse the correspondence, we plot the maximum Faraday angle as a function of $B$ for both the semi-Dirac and Dirac cases as shown in the inset of Fig.~\ref{fig:6}. There is a steady growth of $\theta^{max}$ (corresponding to the peak value of Re($\sigma_{xy}$)) and a decline in $\theta^{min}$ (corresponding to the peak in the negative direction) for both the cases. The behavior observed for the Faraday rotation angle and its dependence on the value  of the magnetic field for the maximum and the minimum rotation angles in graphene\cite{marel}, namely, $\theta^{max}$ and $\theta^{min}$ match well with the Dirac case presented here. 
\section{Conclusion}\label{IV}
We have investigated MO transport properties of a semi-Dirac system subjected to an external magnetic field and compared them to those for the Dirac systems. Owing to the different properties of the Landau levels, the MO conductivities show several distinct features for the semi-Dirac case as compared to the Dirac one. The real parts of the longitudinal conductivities (Re($\sigma_{xx}$) and Re($\sigma_{yy}$)) in either of the cases acquire a series of absorption peaks owing to the transition between the Landau levels with the semi-Dirac case having additional features owing to an asymmetric distribution of the Landau levels and their densities of states. Also, the peak intensity for Re($\sigma_{yy}$) is one order of magnitude larger than that of Re($\sigma_{xx}$) as the semi-Dirac case has relativistic dispersion in the $y$-direction (non-relativistic in the $x$-direction) and hence entails a larger velocity than that in the $x$-direction. Further, in the case of MO Hall conductivity, the semi-Dirac case shows extra absorption peaks for the real, as well as the imaginary parts, owing to the optical transitions. Moreover, we have studied the effect of electron filling on the absorption spectra by tuning chemical potential between the consecutive Landau levels. Also to explore the interplay of the polarization of the incident radiation with MO transport, we consider the light of a different polarization, namely, a circularly polarized light. Particularly, the right circularly polarized beam yields distinct features for the semi-Dirac case compared to the Dirac one. Finally, to ascertain the MO activity of the semi-Dirac systems, we study Faraday rotation where we obtained two discernible Faraday rotation angles for the semi-Dirac case. It should be possible that this feature may be realized in the experiments.
\begin{acknowledgments} 
PS acknowledges Sim$\tilde{a}$o M. Jo$\tilde{a}$o for useful discussions. SM is supported by JSPS KAKENHI under Grant No. JP18H03678.
\end{acknowledgments}
\appendix*
\section{Chebyshev Expansion}
The first kind Chebyshev polynomials can be defined as, $T_n(x)$ = cos($n$ cos$^{-1}(x))$ in the range [$-$1, 1]. The recursion relations, $T_0(x)=1$, $T_1(x)=x$ and $T_{n+1}(x)=2xT_{n}(x)-T_{n-1}(x)$ and the orthogonality relation,
\begin{equation}
\int_{-1}^{1} T_n(x) T_m(x) \frac{dx}{\sqrt{1-x^2}}=\delta_{nm} \frac{1+\delta_{n0}}{2}
\label{eq:14}
\end{equation}
are satisfied by these polynomials.  
The expansion of the Dirac delta in terms of Chebyshev polynomials, can be written as,
\begin{equation}
\delta(\epsilon - H_0)= \sum_{n=0}^{\infty}\Delta_{n}(\epsilon)\frac{T_n(H_0)}{1+\delta_{n0}},
\label{eq:15}
\end{equation}
where
\begin{equation}
\Delta_n(\epsilon)=\frac{2T_n(\epsilon)}{\pi\sqrt{1-\epsilon^2}}.
\label{eq:17}
\end{equation}
Also the Green`s function can be expressed in terms of the $T_n(x)$ as, 
\begin{equation}
\textsl{g}^{\sigma,\lambda}(\epsilon,H_0)=\hbar\sum_{n=0}^{\infty}\textsl{g}^{\sigma,\lambda}_{n}(\epsilon)\frac{T_n(H_0)}{1+\delta_{n0}},
\label{eq:16}
\end{equation}
where
\begin{equation}
\textsl{g}^{\sigma,\lambda}_{n}(\epsilon)=-2\sigma i \frac{e^{-ni\sigma~\mathrm{cos^{-1}}(\epsilon+i\sigma\lambda)}}{\sqrt{1-(\epsilon+i\sigma\lambda)^2}}.
\label{eq:18}
\end{equation}
The function $\textsl{g}^{\sigma,\lambda}$ represents both the retarded and the advanced Green's function in the limit $\lambda \rightarrow 0^+$, where $\lambda$ is the finite broadening parameter. $\textsl{g}^{+,0^+}$ and $\textsl{g}^{-,0^+}$ are the retarded and the advanced Green's function respectively. Hence, the Dirac deltas and Green's function are combinations of a polynomial of $H_0$ (the unperturbed Hamiltonian) and a coefficient which are functions of the frequency and the energy parameters. The trace in the conductivity can be written as a trace over a product of polynomials and $\hat{h}$ operators. The $\Gamma$ matrix needed in the expression of conductivity is written as,
\begin{equation}
\Gamma^{{\boldsymbol{\alpha}_{1}},\dotsb,{\boldsymbol{\alpha}_{m}}}_{n_1\dotsb n_m}=\frac{\mathrm{Tr}}{N}\Bigg[\tilde{h}^{\boldsymbol{\alpha}_{1}}\frac{T_{n_1}(H_0)}{1+\delta_{n_10}}\dotsb\tilde{h}^{\boldsymbol{\alpha}_{m}}\frac{T_{n_m}(H_0)}{1+\delta_{n_m0}}\Bigg], 
\label{eq:19}
\end{equation}
where $N$ is the number of unit cells. The upper indices can be used for any number of indices: ${\boldsymbol{\alpha}_{1}}=\alpha^{1}_1\alpha^{2}_1\dotsb \alpha^{N_1}_1$ and $\tilde{h}^{\boldsymbol{\alpha}_{1}}=(i\hbar)^{N_1}\hat{h}^{\boldsymbol{\alpha}_{1}}$. Here the coefficients of the Chebyshev expansion can be written similarly in a matrix form as,
\begin{equation}
\Lambda_n=\int_{-\infty}^{\infty}d\epsilon f(\epsilon)\Delta_n(\epsilon)
\label{eq:20}
\end{equation}
and 
\begin{align}
\Lambda_{nm}(\omega)=&\hbar\int_{-\infty}^{\infty}d\epsilon f(\epsilon)\big[\textsl{g}^{R}_{n}(\epsilon/\hbar+\omega)\Delta_{m}(\epsilon) \nonumber
\\
&
+\Delta_{n}(\epsilon)\textsl{g}^{A}_{m}(\epsilon/\hbar-\omega)\big],
\label{eq:21}
\end{align}
where $f(\epsilon)=(1+e^{\beta(\epsilon-\mu)})^{-1}$ is the Fermi-Dirac distribution function, where $\beta$ is the inverse temperature and $\mu$ is the chemical potential. 
Hence the first-order conductivity becomes,
\begin{equation}
\sigma^{\alpha\beta}(\omega)=\frac{-i e^2}{\Omega_c \hbar^2 \omega}\Bigg[\sum_{n}\Gamma^{\alpha\beta}_{n}\Lambda_{n}+\sum_{nm}\Lambda_{nm}(\omega)\Gamma^{\alpha,\beta}_{nm}\Bigg]
\label{eq:22}
\end{equation}
where $\Omega_c$ is the volume of the unit cell.

\end{document}